\UseRawInputEncoding
\documentclass[a4paper,11pt]{article}
\pdfoutput=1 
\usepackage{jheppub} 
\usepackage[T1]{fontenc} 
\usepackage{natbib}
\usepackage{hyperref}
\usepackage{tensor}
\usepackage{soul,xcolor}
\usepackage{diagbox}

\def\ba{\begin{eqnarray}}
\def\ea{\end{eqnarray}}
\newcommand{\rmd}{{\rm d}}
\newcommand{\pd}{\partial}
\newcommand{\Lag}{\mathcal{L}}
\newcommand{\beq}{\begin{equation}}
\newcommand{\eeq}{\end{equation}}
\newcommand{\bal}{\begin{aligned}}
\newcommand{\eal}{\end{aligned}}

\title{\boldmath Schwarzschild quasi-normal modes of non-minimally coupled vector fields}

\author[a,b]{Sebastian Garcia-Saenz}
\author[b,c,d]{Aaron Held}
\author[b,e,f]{Jun Zhang}

\affiliation[a]{
	Department of Physics, Southern University of Science and Technology, 
	\\
	Shenzhen 518055, China
}
\affiliation[b]{
	Theoretical Physics, Blackett Laboratory,
	\\
	Imperial College London, SW7 2AZ London, U.K.
}
\affiliation[c]{
Theoretisch-Physikalisches Institut, Friedrich-Schiller-Universit\"at Jena,
\\
Max-Wien-Platz 1, 07743 Jena, Germany}

\affiliation[d]{
The Princeton Gravity Initiative, Jadwin Hall, Princeton University,
\\
Princeton, New Jersey 08544, U.S.}
\affiliation[e]{
	International Centre for Theoretical Physics Asia-Pacific, 
	\\
	Beijing 100190, China
}
\affiliation[f]{Taiji Laboratory for Gravitational Wave Universe, 
\\ 
University of Chinese Academy of Sciences, Beijing 100049, China
}

\emailAdd{sgarciasaenz@sustech.edu.cn}
\emailAdd{aaron.held@uni-jena.de}
\emailAdd{zhangjun@ucas.ac.cn}

\abstract{
	We study perturbations of massive and massless vector fields on a Schwarz\-schild black-hole background, including a non-minimal coupling between the vector field and the curvature. The coupling is given by the Horndeski vector-tensor operator, which we show to be unique, also when the field is massive, provided that the vector has a vanishing background value. 
	
	We determine the quasi-normal mode spectrum of the vector field, focusing on the fundamental mode of monopolar and dipolar perturbations of both even and odd parity, as a function of the mass of the field and the coupling constant controlling the non-minimal interaction. In the massless case, we also provide results for the first two overtones, showing in particular that the isospectrality between even and odd modes is broken by the non-minimal gravitational coupling.
	
	We also consider solutions to the mode equations corresponding to quasi-bound states and static configurations. Our results for quasi-bound states provide strong evidence for the stability of the spectrum, indicating the impossibility of a vectorization mechanism within our set-up. For static solutions, we analytically and numerically derive results for the electromagnetic susceptibilities (the spin-1 analogs of the tidal Love numbers), which we show to be non-zero in the presence of the non-minimal coupling.
}

\begin{document} 
\maketitle
\flushbottom

\section{Introduction} \label{sec:intro}

Black holes (BHs) are arguably among the most interesting objects in the universe. Their experimental agreement with the gravitational wave (GW) emission of binary systems \cite{LIGOScientific:2016aoc} and with the imaging of the event horizon of a supermassive BH \cite{EventHorizonTelescope:2019dse}, promises to be but the first phase of a research era that is bound to culminate in a deeper understanding on the nature of BHs and gravity, as well as on numerous other related questions in astrophysics and fundamental particle physics.

As with most physical systems, a powerful way to probe BHs is by perturbing them and then see how they respond. While we cannot do this in the lab, such perturbed BHs are naturally produced by the merger of compact astrophysical objects. The details of how the post-merger BH relaxes toward equilibrium may in principle be measured through the GWs emitted during the process---the so-called ringdown phase. Quantitatively, the dynamics of the ringdown can be modeled by a superposition of quasi-normal modes (QNMs), whose characteristic frequencies are in one-to-one correspondence with the observable GW signal.

In the context of general relativity (GR), the study of linear perturbations on vacuum and electro-vacuum BH spacetimes has a long history \cite{Regge:1957td,Zerilli:1970se,Zerilli:1970wzz,Moncrief:1974am,Chandrasekhar:1975zza,Teukolsky:1972my,Moncrief:1974gw,Moncrief:1974ng}, and the corresponding QNM spectra are by now well understood; see \cite{Nollert:1999ji,Kokkotas:1999bd,Berti:2009kk,Konoplya:2011qq} for reviews. More recently, QNMs have received renewed interest as valuable indicators of modifications of gravity \cite{Berti:2018vdi}, motivating their analysis in theories beyond GR. For instance, QNMs have been investigated in models of gravity with curvature corrections \cite{Cardoso:2018ptl,Franciolini:2018uyq,deRham:2020ejn,Cano:2021myl} as well as in scalar-tensor theories \cite{Cardoso:2009pk,Molina:2010fb,Tattersall:2018nve,Wagle:2021tam,Bryant:2021xdh,Pierini:2021jxd,Blazquez-Salcedo:2016enn}.

Although radiation in the form of GWs is a universal outcome of perturbing a BH, it need not be the only one. Indeed, a dramatic event such as a BH merger may reasonably be expected to excite other fields besides the metric, and these too will subsequently relax back to equilibrium via emission of the corresponding radiation. This radiation can again be described by QNMs, i.e.\ characterized in particular by dissipation caused by the presence of the BH horizon. More importantly, the matter fields' QNM spectra depend on the underlying spacetime, thus serving as an alternative probe of the BH. Furthermore, and crucially, the QNMs of a field encode physical information that is not directly available in the GW signal, namely about the coupling of the respective field with gravity and the equivalence principle.

In view of these considerations, we see at least two reasons that motivate the study of QNMs of matter fields in a BH background. The first concerns the fields themselves. As we have said, BH mergers are phenomena unlike anything we might achieve with Earth-based experiments. Thus, we may hope to make use of them as a way to test the existence of new particles, especially those that dominantly interact with the Standard-Model sector indirectly through gravity. The second reason which we have already alluded to regards the question of how fields couple to gravity. Establishing the existence of matter-gravity interactions beyond those dictated by the minimal coupling prescription is an exciting prospect that may in principle be achieved through the measurement of QNMs. In fact, as we will discuss later, QNMs offer a particularly clean signature of non-minimal gravitational interactions.

In this paper, we study QNMs of a massive vector field on a Schwarzschild BH background with a particular non-minimal coupling with gravity. Before describing our set-up in detail, let us briefly comment on the existing literature on the subject of vector-field QNMs in BH spacetimes. The study of massless, minimally coupled vector fields in four dimensions and with flat asymptotics dates back to the work of Chandrasekhar \cite{Chandrasekhar:1985kt}. The Proca equation for a massive vector field and the corresponding QNM spectra have been investigated in \cite{Galtsov:1984ixy,Konoplya:2005hr,Rosa:2011my,Fernandes:2021qvr} for a Schwarzschild(-AdS) BH and only recently in \cite{Frolov:2018ezx,Baumann:2019eav,Percival:2020skc} for a Kerr BH.

Here we go beyond previous studies of spin-1 particles by considering the most general Lagrangian of a vector field $A_{\mu}$ subject to the following assumptions:
\begin{itemize}
\item[(i)] The Lagrangian is quadratic in the vector field. This follows from our aim to investigate linear perturbations about vacuum solutions of general relativity (GR), specifically the Schwarzschild metric. Generically, this implies that the vector field must vanish at the background level, and therefore it is sufficient to focus on a quadratic theory for the purpose of deriving the QNM spectrum.

While this is true generically, we should remark that there exist vector-tensor theories that admit so-called ``stealth'' BH solutions, i.e.\ solutions that coincide with vacuum GR solutions in spite of having a non-trivial vector field background profile \cite{Chagoya:2017ojn}. Our analysis therefore does not encompass this case.

A corollary of this premise is that metric and vector perturbations are decoupled at linear order. The QNM spectrum of GWs is thus exactly the one derived in GR \cite{Chandrasekhar:1975zza,Chandrasekhar:1985kt} and so may be ignored.

\item[(ii)] The theory describes precisely five dynamical degrees of freedom, i.e.\ two in the metric and three in the vector field (or two in the case of a massless spin-1 field, which we will treat as a special case). In other words, we demand the absence of additional propagating modes associated to a loss of constraints or to higher-order equations of motion.

\end{itemize}

For a generic spacetime background, this Lagrangian extends the Proca theory by the addition of two non-minimal coupling operators. Unsurprisingly, these operators are precisely those obtained from linearizing the Lagrangian of the Generalized Proca theory of a self-interacting massive spin-1 field \cite{Tasinato:2014eka,Heisenberg:2014rta}. Our derivation thus serves as a proof of the uniqueness of Generalized Proca theory at the level of linear perturbations about the trivial state $\langle A_{\mu}\rangle=0$. For a Ricci-flat background the theory further simplifies, leaving only one non-minimal coupling operator,
\beq \label{eq:g6 coupling}
\Lag\supset \sqrt{-g}\,G_6R^{\mu\nu\rho\sigma}F_{\mu\nu}F_{\rho\sigma} \,,
\eeq
with $F_{\mu\nu}$ the vector field strength and $G_6$ a coupling constant.

Our main objective in this paper is to numerically derive the QNM spectrum of vector field perturbations for a range of values of the parameter $G_6$ and the bare mass $\mu$ of the field. Interestingly, for a given BH mass, $G_6$ is restricted to a window of values given by
\beq \label{eq:bound g6}
-\frac{r_g^2}{2}<G_6<r_g^2 \,,
\eeq
where $r_g\equiv 2GM$ is the Schwarzschild radius (with $G$ the Newton coupling and $M$ the BH mass). This criterion follows from the requirement of stability (due to ghost and/or gradient instabilities) of the BH under perturbations of the vector field in the localized approximation, i.e.\ in the limit where the size of the perturbation is much shorter than the typical length scale characterizing the background variation \cite{Jimenez:2013qsa,Garcia-Saenz:2021uyv}.

Related to the question of stability of BH spacetimes under perturbations of generalized vector fields, one may ask if the criterion \eqref{eq:bound g6} is not only necessary but also sufficient for ensuring stability. While we plan to address this with exhaustivity in a dedicated work, here we provide evidence that this is indeed the case for a Schwarzschild BH. Our claim is based on the analysis of quasi-bound states of the vector field, that is solutions of the generalized Proca equation which decay at spatial infinity. Like QNMs, quasi-bound states have an associated spectrum of complex frequencies, which are of particular interest as they may be used to diagnose the presence of instabilities: A quasi-bound state frequency with positive imaginary part signals an exponentially growing mode and thus an unstable system, at least within the linearized regime. It is worth remarking that the same judgment cannot be made based on the QNM spectrum because, as we will review, the imaginary part of a QNM frequency must be negative by the definition of a QNM.

The principal result of this first study of quasi-bound states of a non-minimally coupled vector field is that the fundamental frequency mode for each degree of freedom of the field has a negative imaginary part within the numerically accessible part of the range given by Eq.\ \eqref{eq:bound g6}. However, as the computational cost of our numerical routine increases as one approaches the bounds in \eqref{eq:bound g6}, we are unable to numerically access values of the coupling $G_6$ arbitrarily close to the critical points. We partially address this shortcoming by providing an analytical argument, valid for a subset of the spectrum, which shows that quasi-bound states are stable whenever $G_6$ is within but arbitrarily close to the stability bounds.

We have mentioned that the QNM spectrum of matter fields may serve as a powerful tool to test the minimal coupling paradigm dictating the form of matter-gravity interactions. This question is of fundamental importance, so it behooves us to understand which signatures of a non-minimal coupling operator for a given field might be clean and robust enough so as to be potentially detectable. Perturbations of a {\it massless} field are arguably one such probe, since minimally coupled massless fields on BH spacetimes in GR are known to be very special, at least due to two properties: isospectrality of their QNM spectra and vanishing linear response coefficients in the static limit.

Isospectrality refers to the equivalence of the QNM spectra of parity-even and parity-odd perturbations \cite{Chandrasekhar:1975nkd,Chandrasekhar:1985kt}. The property is featured by massless fields in four dimensions with flat or de Sitter asymptotics, at least for spins $s=0,1,2$.\footnote{In the case of a Schwarzschild-de Sitter BH, isospectrality also holds for partially massless spin-2 perturbations \cite{Brito:2013yxa,Rosen:2020crj}.} Isospectrality is however known to fail in higher dimensions \cite{Konoplya:2003dd}, for asymptotically anti-de Sitter BHs \cite{Cardoso:2001bb,Berti:2009kk}, for massive fields \cite{Rosa:2011my,Brito:2013wya}, and for BHs in non-linear electrodynamics \cite{Chaverra:2016ttw,Nomura:2021efi} or in the presence of higher-curvature corrections \cite{Cardoso:2018ptl,deRham:2020ejn}. To our knowledge, the breaking of isospectrality due to non-minimal couplings has not been systematically addressed, although it is known to occur for certain couplings of scalar fields \cite{Molina:2010fb,Wagle:2021tam,Blazquez-Salcedo:2016enn,Pierini:2021jxd,Bryant:2021xdh}. Here we fill the gap of spin $s=1$ by showing through numerical results that the parity-even and -odd spectra of a massless vector field with the non-minimal coupling of eq.\ \eqref{eq:g6 coupling} are indeed distinct.

Although QNMs are the main focus of our work, {\it static} perturbations are also interesting in that they define the static response coefficients associated to a given field. For massless spin-2 perturbations the response coefficients physically encode the tidal deformability of the BH and are known as Love numbers \cite{Love1912}, see also~\cite{Flanagan:2007ix,Damour:2009vw,Binnington:2009bb}. For a massless spin-1 probe field they may be interpreted as the electromagnetic susceptibilities of the field in a BH background. It is a remarkable and well-known property that the static response coefficients of massless fields of spin $s=0,1,2$ exactly vanish for four-dimensional BHs in GR \cite{LeTiec:2020spy,Chia:2020yla,Goldberger:2020fot,Hui:2020xxx,LeTiec:2020bos,Charalambous:2021mea,Pereniguez:2021xcj}. The property is however absent in higher dimensions \cite{Kol:2011vg,Cardoso:2019vof} as well as for BHs in beyond-GR theories \cite{Cardoso:2017cfl,Cardoso:2018ptl,Cai:2019npx}. In addition, and similarly to isospectrality, the vanishing of Love numbers and susceptibility coefficients is not expected to hold in the presence of non-minimal couplings, although again we are not aware of any exhaustive analyses (see \cite{Cardoso:2017cfl} for results in some particular models). Here, we compute the electric and magnetic susceptibilities of dipolar perturbations of a massless vector field as functions of the coupling $G_6$ in eq.\ \eqref{eq:g6 coupling}, and show that they are non-vanishing in agreement with expectations.

We now give an outline of the paper's contents: In Sec.\ \ref{sec:setup}, we describe our set-up, including (i) our uniqueness argument for the non-minimal coupling, (ii) the decomposition of the vector field in spherical harmonics, and (iii) the definition of QNMs according to the boundary conditions for the mode equations. In Sec.\ \ref{sec:qnm}, we present our main results, namely the calculation of the QNM spectra for each mode of the vector field and for a range of values of the coupling $G_6$ and mass $\mu$. The spectra of a massless field and the breaking of isospectrality are treated as a special case. In Sec.\ \ref{sec:bound states}, we consider quasi-bound states. This provides evidence for the stability of the system under consideration beyond the local approximation. In Sec.\ \ref{sec:susceptibilities}, we consider static perturbations and, focusing on a massless field and dipolar modes, derive the electric and magnetic susceptibilities as functions of $G_6$. We discuss our results and give some final remarks in Sec.\ \ref{sec:discussion}. In Appendix \ref{sec:appendix}, we provide details of the numerical method used in our calculations.


\section{Non-minimally coupled Proca field}
\label{sec:setup}

Our study will focus on linear perturbations of a massive vector field $A_{\mu}$ about a GR background solution. The background state of the vector field is the trivial one, $\langle A_{\mu}\rangle=0$, as per our definition of a GR solution, i.e.\ one with vanishing vector hair. It therefore suffices to focus on Lagrangians that are precisely quadratic in the field $A_{\mu}$, while the dependence on the metric tensor is in principle arbitrary. Note that, a priori, we make no restriction on the number of derivatives acting on $A_{\mu}$.

We will additionally require that the theory describe exactly five degrees of freedom---two in the metric and three in the vector field---so as to avoid Ostrogradsky-type ghosts on all backgrounds. A sufficient condition to achieve this is to demand that the equations of motion of the St\"uckelberg formulation of the theory be of second order in derivatives. This condition is however not a priori necessary, as it may occur that the theory possess the correct number of constraints even in the presence of higher derivatives in the field equations \cite{deRham:2016wji}. We shall nevertheless disregard this possibility here and focus on the simpler set-up with second-order equations of motion.

Our claim is that the most general four-dimensional Lagrangian subject to these assumptions is given by
\beq\bal \label{eq:full lagrangian}
\mathcal{L}&=\sqrt{-g}\bigg[\frac{M_{\rm Pl}^2}{2}\,R-\frac{1}{4}\,F^{\mu\nu}F_{\mu\nu}-\frac{\mu^2}{2}\,A^\mu A_\mu+G_{4,X}A^{\mu}A^{\nu}G_{\mu\nu} \\
&\quad -\frac{G_6}{4}\left(F^{\mu\nu}F_{\mu\nu}R-4F^{\mu\rho}F^{\nu}_{\phantom{\nu}\rho}R_{\mu\nu}+F^{\mu\nu}F^{\rho\sigma}R_{\mu\nu\rho\sigma}\right)\bigg] \,,
\eal\eeq
where $M_{\rm Pl}$ is the Planck mass, $\mu$ is the mass of the vector field, and $G_{4,X}$ and $G_6$ are coupling constants. The notation chosen for the latter two coefficients is explained by the connection between the Lagrangian \eqref{eq:full lagrangian} and the Generalized Proca theory. As mentioned in the introduction, eq.\ \eqref{eq:full lagrangian} may be obtained upon linearizing the Generalized Proca Lagrangian about the trivial vector background $\langle A_{\mu}\rangle=0$.\footnote{The Generalized Proca Lagrangian contains the functions $G_4(X)$ and $G_6(X)$ (among others), with $X\equiv -\frac{1}{2}\,A^{\mu}A_{\mu}$. Our coupling constants $G_{4,X}$ and $G_6$ correspond respectively to $G_4'(0)$ and $G_6(0)$, which are finite by our assumption that $\langle A_{\mu}\rangle=0$ is a well-defined state.} In particular, the operators multiplying $G_6$ in the second line (which may be more compactly written in terms of the dual Riemann tensor) will be recognized as the unique extension, as demonstrated by Horndeski \cite{Horndeski:1976gi}, of the standard Einstein-Maxwell theory, here restricted to quadratic order.

In this paper we confine our attention to a background given by the Schwarzschild metric and neglect the backreaction of the vector field on the geometry. This assumption is valid at linear order in perturbation theory since, as we remarked, metric and vector fluctuations do not couple at this order. The generalized Proca equation for a Ricci-flat spacetime reduces to
\beq \label{eq:covariant eom}
\nabla_{\mu}F^{\mu\nu}+G_6R^{\mu\nu\rho\sigma}\nabla_{\mu}F_{\rho\sigma}-\mu^2A^{\nu}=0 \,.
\eeq
In this set-up, we are therefore left with two dimensionless parameters: $\mu r_g$ and $g_6\equiv G_6/r_g^2$ (with $r_g$ the Schwarzschild radius). Observe that the Lorenz constraint,
\beq \label{eq:lorenz condition}
\nabla_{\mu}A^{\mu}=0 \,,
\eeq
follows as a consequence of eq.~\eqref{eq:covariant eom} whenever $\mu\neq0$. In the massless case, we shall instead impose a different constraint as a gauge condition.

As mentioned in the introduction, eq.\ \eqref{eq:covariant eom} features pathological solutions (ghosts and/or gradient-unstable modes) unless the coefficient $g_6$ is confined to the range \cite{Jimenez:2013qsa,Garcia-Saenz:2021uyv}
\beq \label{eq:bounds g6}
-\frac{1}{2}<g_6<1 \,.
\eeq
While this result was obtained from an analysis of localized perturbations, these bounds on $g_6$ will be seen to translate into the statement that the mode functions of the vector field should be insensitive to additional poles appearing in the equations of motion. In terms of the Schwarzschild radial coordinate $r$, these poles are given by
\beq \label{eq:poles}
P_{\pm}\equiv 1-\frac{r_{\pm}^3}{r^3} \,,
\eeq
with $r_{+}\equiv g_6^{1/3}r_g$ and $r_{-}\equiv (-2g_6)^{1/3}r_g$. Demanding that these poles be hidden inside the event horizon then yields \eqref{eq:bounds g6}. Thus, although this range was originally derived from different considerations, it has the important implication that the equations will allow for consistent QNM solutions, which at least generically would not be possible if one had poles in the physical domain $r>r_g$.\footnote{QNMs are by definition everywhere regular and with fixed boundary conditions. The presence of a pole would impose an additional matching condition and thus an overdetermined system for the QNM frequency and the amplitude of the QNM function. Such a system will generically have no solution. The same remark, of course, also applies to quasi-bound states.}

\subsection{Uniqueness}
\label{sec:unique}

Generalized Proca theory was constructed as the most general model which reproduces the (shift-symmetric) scalar Horndeski theory in the so-called decoupling limit where the longitudinal mode of the vector field becomes a dynamical scalar \cite{Allys:2015sht,BeltranJimenez:2016rff}.\footnote{Other prescriptions for constructing vector-tensor theories have been considered in the literature \cite{Heisenberg:2016eld,Kimura:2016rzw,deRham:2020yet,deRham:2021efp}, leading to various extensions of Generalized Proca. See also \cite{Aoki:2021wew} for an effective field theory approach.} As such, the theory is perfectly general, given the assumptions of its construction, on flat spacetime. The uniqueness of Generalized Proca is however not immediate when the coupling with gravity is taken into account, since the covariantization of the decoupling limit theory need not match, term by term, that of the full theory. In particular, one cannot a priori disregard non-minimal couplings to the curvature tensor beyond those obtained in \cite{Heisenberg:2014rta} (see also \cite{Hull:2015uwa}), as the latter were derived as ``counterterms'' to cancel the pathological operators that appear upon minimal covariantization. Here, we provide a sketch of the proof of the uniqueness of the Lagrangian \eqref{eq:full lagrangian}; a detailed proof will be given in a dedicated work where we analyze the general problem without assuming linearity in the vector field.

To reiterate the problem, we seek the most general Lagrangian for a vector field $A_{\mu}$ and metric tensor $g_{\mu\nu}$ subject to the assumptions of (i) general covariance, (ii) quadratic order in the vector field, and (iii) second-order equations of motion for all the fields in the St\"uckelberg formulation. Note that we make no assumption on the derivative order of the fields at the level of the Lagrangian.

In the St\"uckelberg formulation of the theory, the Lagrangian is a functional of the fields $(g_{\mu\nu},A_{\mu},\phi)$ and is invariant under diffeomorphism and $U(1)$ gauge symmetries. The latter property implies that the vector and St\"uckelberg scalar can only appear through the invariants $F_{\mu\nu}$ and ${D}_{\mu}\phi\equiv \nabla_{\mu}\phi+\mu A_{\mu}$ (here $\mu$ is the mass of the Proca field). The main proposition is that these two building blocks cannot couple with each other in the Lagrangian. To establish this one notes that covariant derivatives of ${D}_{\mu}\phi$ may be chosen as fully symmetrized without loss of generality. Indeed, any mixed-symmetric or antisymmetric projection of $\nabla_{\mu_1}\cdots\nabla_{\mu_{n-1}}{D}_{\mu_n}\phi$ can be traded by $F_{\mu\nu}$ (and derivatives thereof) and/or curvature tensors contracted with fully symmetrized derivatives of ${D}_{\mu}\phi$. Since derivatives of $F_{\mu\nu}$ cannot be made fully symmetric, it follows that they cannot be contracted with the tensor $\nabla_{(\mu_1}\cdots\nabla_{\mu_{n-1}}{D}_{\mu_n)}\phi$. An exception to this is the divergence of the field strength, $\nabla_{\mu}F^{\mu\nu}$, and its derivatives; for instance, $\nabla_{\mu}F^{\mu\nu}{D}_{\nu}\phi$ is a valid operator that seemingly contradicts our claim. However, the divergence $\nabla_{\mu}F^{\mu\nu}$ may in principle be solved for algebraically from the vector field equation of motion, implying that any instance of this term in the Lagrangian may be eliminated through a field redefinition.

The Lagrangian is therefore ``separable'' in the building blocks $F_{\mu\nu}$ and ${D}_{\mu}\phi$, which may then be analyzed independently. The operators involving only $F_{\mu\nu}$ and the metric constitute purely vector-tensor gauge invariant terms, hence they satisfy the assumptions of the Horndeski theorem for Einstein-Maxwell theory \cite{Horndeski:1976gi}, with the known quadratic-order result\footnote{The double-dual Riemann tensor is defined as $\widetilde{R}^{\mu\nu\rho\sigma}\equiv \frac{1}{4}\,\epsilon^{\mu\nu\mu'\nu'}\epsilon^{\rho\sigma\rho'\sigma'}R_{\mu'\nu'\rho'\sigma'}$.}
\beq
\Lag\supset \frac{1}{4}\,\sqrt{-g}\,G_6\widetilde{R}^{\mu\nu\rho\sigma}F_{\mu\nu}F_{\rho\sigma} \,,
\eeq
together with the standard Maxwell, Einstein-Hilbert and cosmological constant terms. The remaining operators in the Lagrangian must then all be expressible in terms of ${D}_{\mu}\phi$ and covariant derivatives of this invariant. Because the conditions we are imposing on the equations of motion must hold for all field configurations, they must hold, in particular, when $A_{\mu}=0$. But in this case ${D}_{\mu}\phi\to \nabla_{\mu}\phi$ and we have precisely the assumptions of the Horndeski theorem for scalar-tensor theory \cite{Horndeski:1974wa} (with the extra condition that $\phi$ may not appear without derivative), with the known quadratic-order result
\beq
\Lag\supset \sqrt{-g}\,\frac{G_{4,X}}{\mu^2}\, G^{\mu\nu}\nabla_{\mu}\phi\nabla_{\nu}\phi \,,
\eeq
together with the standard scalar kinetic term. In the general case with $A_{\mu}\neq 0$, we know that the scalar field derivative must appear ``covariantized'' in ${D}_{\mu}\phi$, so that the result correctly reproduces the Generalized Proca term upon setting unitary gauge $\phi=0$.

This concludes our derivation of the Lagrangian \eqref{eq:full lagrangian}, independently of its relation with the non-linear Generalized Proca theory. The implication is that any consistent extension or alternative to Generalized Proca must reduce to \eqref{eq:full lagrangian} when expanded at quadratic order about the vacuum $\langle A_{\mu}\rangle=0$, provided the theory admits this state.

\subsection{Decomposition in vector spherical harmonics}
\label{sec:decomp}

We consider the exterior of a Schwarzschild BH spacetime with line element
\begin{align}
	ds^2 = - f(r) dt^2 + \frac{1}{f(r)} dr^2 + r^2 (d\theta^2+\sin^2\theta\,d\phi^2)\;,
	\quad\quad\text{with}\quad
	f(r) = 1-\frac{r_g}{r}\;,
\end{align}
where $r_g=2GM$, $G$ is the Newton coupling, and $M$ is the mass of the BH. Given the background symmetries, the equation of motion for the vector field is separable after expanding in spherical harmonics,
\begin{align}
\label{eq:ansatz}
	A_{\mu}(t,r,\theta,\phi) = \frac{1}{r} \sum_{i=1}^{4} \sum_{\ell,\,m} u^{\ell m}_{i}(t,r) Z_\mu^{(i)\ell m}(\theta, \phi) \;,
\end{align}
where, in our convention, the vector spherical harmonics are defined as
\begin{align} 
\label{eq:vector-harmonics}
Z_{\mu}^{(1)\ell m} &= \left[f(r),\,0,\,0,\,0\right] Y^{\ell m}\;, \\
Z_{\mu}^{(2)\ell m} &= \left[0,\,f(r)^{-1},\,0,\,0\right] Y^{\ell m}\;, \\
Z_{\mu}^{(3)\ell m} &= \frac{r}{\ell (\ell +1)}\left[0,\,0,\,\partial_\theta,\,\partial_\phi\right] Y^{\ell m}\;, \\
Z_{\mu}^{(4)\ell m} &= \frac{r}{\ell (\ell +1)}\left[0,\,0,\,\csc\theta\,\partial_\phi,\,-\sin\theta\,\partial_\theta\right] Y^{\ell m}\;,
\end{align}
in terms of the standard scalar spherical harmonics $Y^{\ell m}(\theta,\phi)$. Under a parity transformation, $\left(\theta \rightarrow \pi-\theta, \, \phi \rightarrow \pi+\phi\right)$, the functions $Z_{\mu}^{(1,2,3)\ell m}$ are even, i.e.\ they pick up a factor $(-1)^\ell$ and the corresponding modes are called polar; $Z_{\mu}^{(4)\ell m}$ is instead odd under parity, transforming with the sign $(-1)^{\ell+1}$, and the modes are called axial. As the background and field dynamics are parity invariant, polar and axial modes are decoupled at linear order and may be analyzed separately. The vector spherical harmonics satisfy the orthonormality condition
\ba
\int d\Omega \, Z_{\mu}^{*(i)\ell m}M_Z^{\mu\nu} Z_{\nu}^{(j)\ell' m'} = \delta_{ij}\delta_{\ell \ell'} \delta_{mm'} \,,
\ea
where $M_Z^{\mu\nu} = {\rm diag}\left[ 1/f^2, f^2, \ell(\ell+1)/r^2, \ell(\ell+1)/(r^2\sin^2\theta) \right]$, which is used to factor out the angular dependence in the equation of motion.

In the following, we suppress the supersripts $\ell$ and $m$ in the mode functions $u_i^{\ell m}$ and denote partial derivatives with respect to $t$ and $r$ respectively with dots and primes. We also introduce the operator
\ba
{\cal D} \equiv -\frac{\pd^2}{\pd t^2} + \frac{\pd^2}{\pd r_*^2}\, ,
\ea
with $r_*$ being the tortoise coordinate defined by $d r_* = f^{-1} dr$. In the next subsection we provide the mode equations for the dynamical degrees of freedom. The case of a massless vector field requires a separate analysis, which is done in the following subsection. Readers interested only in the relevant equations and results may consult Tab.~\ref{tab:perturbations-overview} for reference.

\begin{table}
	\begin{center}
		\begin{tabular}{r||l|l|l||l|l} 
			& function
			& equation
			& poles
			& QNMs
			& QBSs
			\\
			\hline\hline
			monopole  &
			$u_M$ &
			\eqref{eq:EoMmono} &
			$\left[P_-\right]$ &
			Fig.~\ref{fig:mono_qnm_n0} &
			Fig.~\ref{fig:boundStates_complex-plane}, \ref{fig:boundStates_powerlaw} 
			\\
			\hline
			axial modes &
			$u_-$ &
			\eqref{eq:EoMaxial} &
			$\left[P_+\right]$ &
			Fig.~\ref{fig:axial_qnm_l1_n0} &
			Fig.~\ref{fig:boundStates_complex-plane}, \ref{fig:boundStates_powerlaw}
			\\ 
			\hline
			polar modes &
			$u_2$, $u_3$ &
			\eqref{eq:EoMpolar2}, \eqref{eq:EoMpolar3} &
			$\left[P_+, P_-\right]$ &
			Fig.~\ref{fig:polar_scalar_qnm_l1_n0} (scalar) &
			Fig.~\ref{fig:boundStates_complex-plane}, \ref{fig:boundStates_powerlaw}
			\\
			&
			$\overset{\mu\rightarrow0}{\longrightarrow}$ $u_0$ &
			\eqref{eq:EoMpolarmassless} &
			 &
			Fig.~\ref{fig:polar_vector_qnm_l1_n0} (vector) & 
			\\ 
			\hline
		\end{tabular}
	\end{center}
	\caption{
		\label{tab:perturbations-overview}
		Reference table for the generalized Proca field mode functions and equations, along with the plots of the results for the QNM and quasi-bound state (QBS) spectra. We also indicate which of the poles in \eqref{eq:poles} feature in each mode equation.
	}
\end{table}

\subsection{Mode equations}
\label{sec:modes}

In order to eliminate the non-dynamical variables we make use of the Lorenz constraint, eq.\ \eqref{eq:lorenz condition}, which reduces to
\begin{align} \label{eq:constraint-explicit}
	\dot{u}_{1} = \frac{f}{r}\left(ru'_{2}+u_{2}-u_{3}\right) \,,
\end{align}
upon substituting the expansion in eq.~\eqref{eq:ansatz}. Here, and in the following, we denote $t$-derivatives with a dot and $r$-derivatives with a prime.

In the case of monopole $(\ell=0)$ perturbations, $u_{3}$ and $u_{4}$ are absent in the spherical harmonic expansion. Using constraint~\eqref{eq:constraint-explicit} we can further eliminate $u_1$ in favor of $u_2\equiv u_M$, with the resulting equation
\ba\label{eq:EoMmono}
{\cal D}u_M - \frac{f}{P_-}\left[\mu^2 + P_-\left(\frac{2}{r^2}-\frac{3 r_g}{r^3}\right)\right] u_M =0 \,.
\ea

For axial perturbations with $\ell \geq 1$, it is convenient to define $u_- \equiv P_+^{1/2} u_4$, which produces
\beq \label{eq:EoMaxial}
{\cal D}u_-  -  \frac{f}{P_+^2} \left[
		\mu^2 P_+
		+ \frac{\ell(\ell+1)}{r^2}\,P_+P_-
		+ \frac{9}{4r^2}\left(P_+ -1\right)\left(
			f+\left(\frac{5}{3}-\frac{7r_g}{3r}\right) P_+
		\right)
	\right]u_- =0 \,.
\eeq
For the polar modes with $\ell \geq 1$ we again eliminate $u_1$ using the constraint \eqref{eq:constraint-explicit}, obtaining two coupled equations for the variables $u_2$ and $u_3$,
\begin{align}
&{\cal D}u_2 -   \frac{f}{P_-}\,\mathcal{V}_2 =0 \,,
\label{eq:EoMpolar2}\\
&{\cal D}u_3 -   \frac{f}{P_+}\,\mathcal{V}_3 =0 \,,
\label{eq:EoMpolar3}
\end{align}
with
\begin{align}
	\mathcal{V}_2 =& 
	\left[
		\mu^2 
		+\frac{\ell(\ell+1)}{r^2}\,P_+
		+\frac{2}{r^2}\left(1-\frac{3r_g}{2r}\right)P_-
	\right] u_2
	-\frac{2}{r^2}\left(1-\frac{3r_g}{2r}\right) P_-u_3
	\notag\\&
	- \left(1-P_-\right)\frac{3f}{2r}\,u_3' \,,
	\\[0.5em]
	\mathcal{V}_3 =& 
	\left[\mu^2 + \frac{\ell(\ell+1)}{r^2}\,P_+\right] u_3 
	+ \frac{\ell(\ell+1)}{r^2}\left(3-5P_+\right)u_2 
	-\left(1-P_+\right)\frac{3f}{r}u_3' \,.
\end{align}
As anticipated in table \ref{tab:perturbations-overview}, the monopole mode is only sensitive to the pole $P_-$, axial modes are only sensitive to the pole $P_+$, while polar modes with $\ell\geq1$ are affected by both. We remind the readers that the parameter $g_6$ (implicit in the above equations, cf.\ \eqref{eq:poles}) is restricted to lie in the stability range \eqref{eq:bounds g6}, so that the poles $P_{\pm}$ never vanish in the physical domain $r> r_g$. Nevertheless, the observation is pertinent as we shall be interested in exploring values of $g_6$ close to the bounds.

\subsection{Massless case} \label{sec:massless}

When the bare mass $\mu$ vanishes, the Lagrangian \eqref{eq:full lagrangian} is gauge invariant and the identification of the dynamical degrees of freedom requires a separate analysis. We will use the gauge freedom to set $u_1=0$ as was done in Ref.~\cite{Rosa:2011my}. Note that this is a complete gauge fixing for perturbations compactly supported in space and time.

For $\ell = 0$, both $u_3$ and $u_4$ are again absent, while the generalized Proca equation implies that $u_2=0$, indicating as expected that there is no dynamical monopole mode. For the higher multipoles with $\ell \geq 1$ we introduce
\begin{align}
u_0 \equiv f\sqrt{P_+} \left(u_3' - \frac{\ell(\ell+1)}{r-r_g} u_2\right) \,,
\end{align}
in terms of which the parity-even part of the equation of motion can be cast as
\begin{align}
\label{eq:EoMpolarmassless}
	{\cal D}u_0 -  \frac{f}{P_+^2  P_-} \left[
	\frac{\ell(\ell+1)}{r^2}\,P_+^3
	+\frac{3}{4r^2}P_-\left(1-P_+\right)\left(
		\left(9-P_+\right)
		-\frac{r_g}{r}\left(9+P_+\right)
	\right)
\right]u_0 = 0 \,,
\end{align}
so that there is a single polar mode (for each $\ell,m$) in the massless case. As for the axial mode, being gauge invariant, one can directly set $\mu=0$ in eq.~\eqref{eq:EoMaxial} to obtain the corresponding equation.

\subsection{Boundary conditions} 
\label{sec:bc}

We seek solutions to the mode equations in the frequency domain, where they assume the form
\begin{align} \label{eq:generic mode eq}
\frac{\rmd^2 u}{\rmd r_*^2} + \omega^2 u -  f \mathcal{V}(u,r) = 0 \,,
\end{align}
for the modes $u_M$, $u_-$, $u_{2,3}$ and $u_0$ (cf.\ table \ref{tab:perturbations-overview}), and remembering that the functional $\mathcal{V}$ couples both modes $u_{2,3}$ in the polar sector. The frequency $\omega$ is in general complex, assuming without loss of generality a positive real part. (If $u$ solves the mode equation for some frequency with ${\rm Re}\,\omega>0$, then $u^*$ solves the conjugate equation with ${\rm Re}\,\omega<0$.) Presently, we further assume ${\rm Im}\,\omega<0$, deferring a discussion of the opposite case to Sec.\ \ref{sec:bound states}.

The BH horizon serves as a causal boundary admitting only ingoing modes, hence the physical boundary condition is
\begin{align}
	u(\omega, r) \sim c_{\rm hor}e^{-i \omega r_*} \,,
\end{align}
at the horizon, i.e.\ as $r_* \rightarrow -\infty$.

At spatial infinity, $r_* \rightarrow +\infty$, we have $\mathcal{V}(u,r) \simeq \mu^2 u$ for every mode. The general asymptotic solution at spatial infinity is therefore
\begin{align} \label{eq:asymptotic sol}
	u(\omega, \,r) \sim c_{\rm out} e^{\sqrt{\mu^2-\omega^2}\,r_*}+c_{\rm in} e^{-\sqrt{\mu^2-\omega^2}\,r_*} \,.
\end{align} 
By definition, QNM solutions correspond to purely outgoing waves at infinity, i.e.\ with $c_{\rm in}=0$.\footnote{To see explicitly that $e^{\sqrt{\mu^2-\omega^2}\,r_*}$ is an outgoing wave, note that we choose the convention for the square root such that ${\rm Re}\sqrt{\mu^2-\omega^2}>0$, which implies that ${\rm sign}({\rm Im}\sqrt{\mu^2-\omega^2})=-{\rm sign}({\rm Im}\,\omega)=+1$.}

Having fixed boundary conditions at both the event horizon and at spatial infinity, we are left with an eigenvalue problem with a discrete set of QNM solutions characterized by a spectrum of frequencies $\{\omega_n\}_{n=0}^{\infty}\,$.


\section{Quasi-normal modes: numerical results} \label{sec:qnm}

\begin{figure}
  \includegraphics[width=\textwidth]{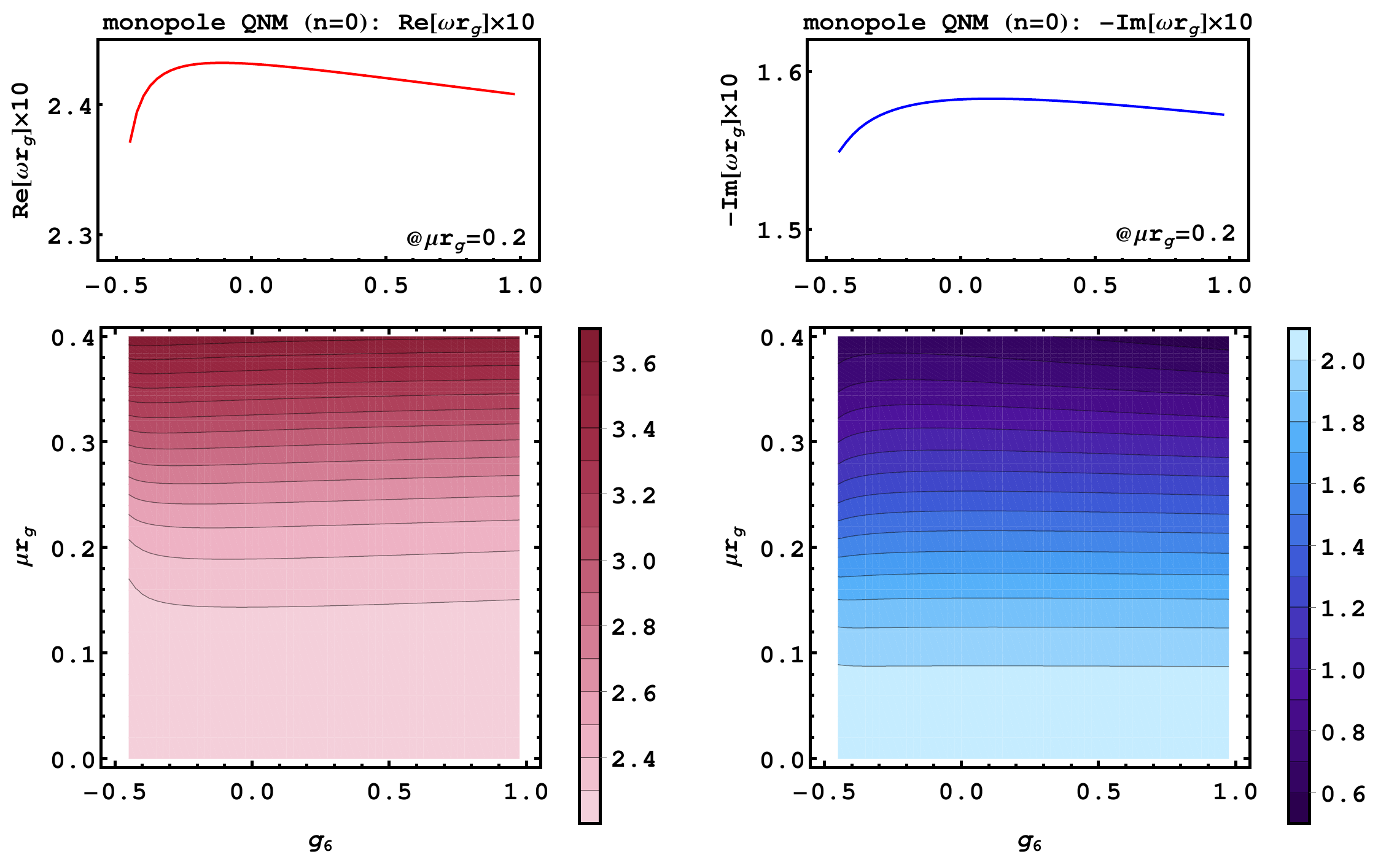}
  \caption{Real (left-hand panels) and imaginary (right-hand panels) part of the fundamental ($n=0$) monopole QNM in the $g_6$--$\mu$ plane (lower panels) and as a function of $g_6$ at $\mu\,r_g=0.2$.}
  \label{fig:mono_qnm_n0}
\end{figure}
\begin{figure}
  \includegraphics[width=\textwidth]{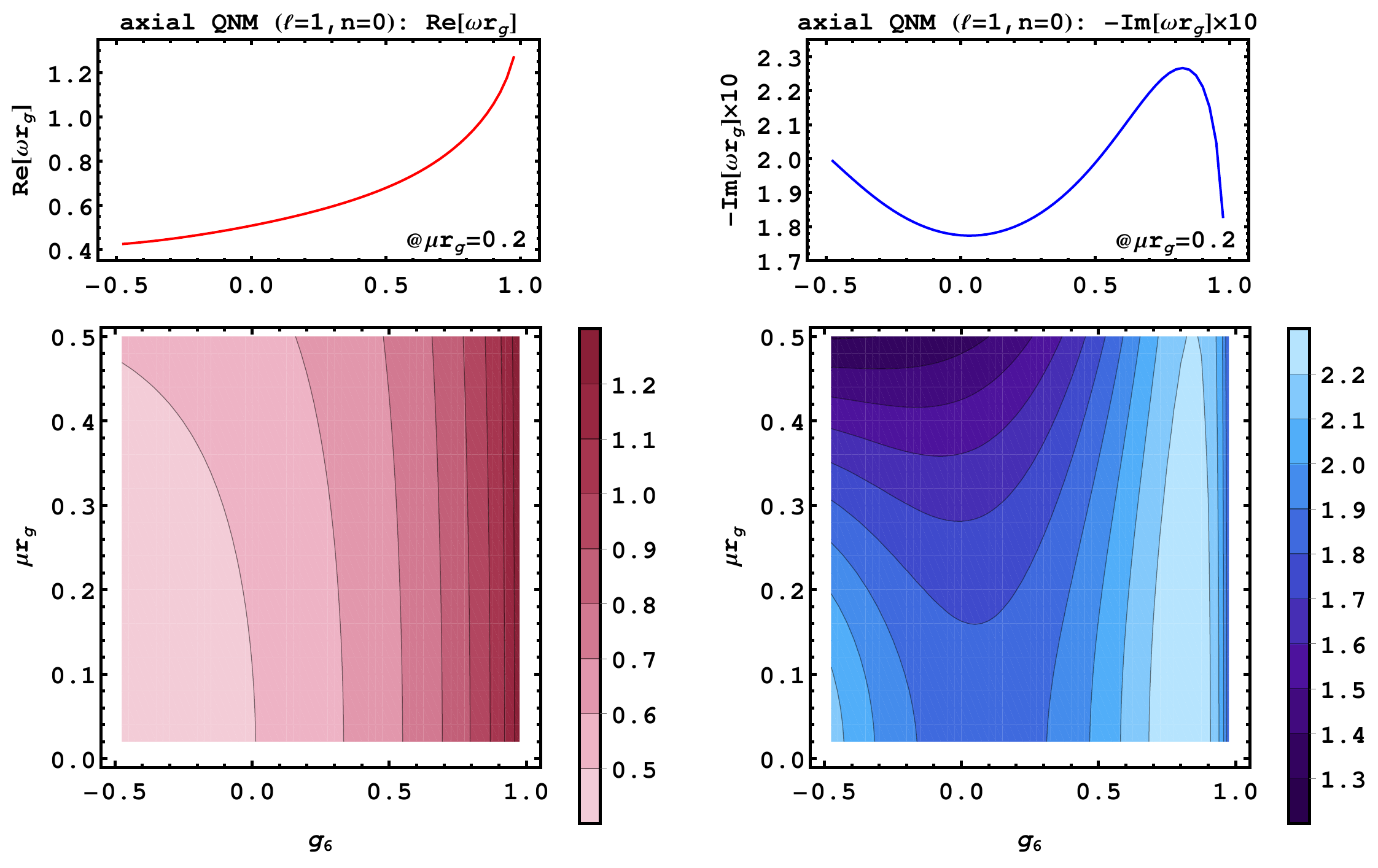}
  \caption{Axial: Real (left-hand panels) and imaginary (right-hand panels) part of the fundamental ($n=0$) first multipole ($\ell=1$) QNM in the $g_6$--$\mu$ plane (lower panels) and as a function of $g_6$ at $\mu\,r_g=0.2$.}
  \label{fig:axial_qnm_l1_n0}
\end{figure}
\begin{figure}
  \includegraphics[width=\textwidth]{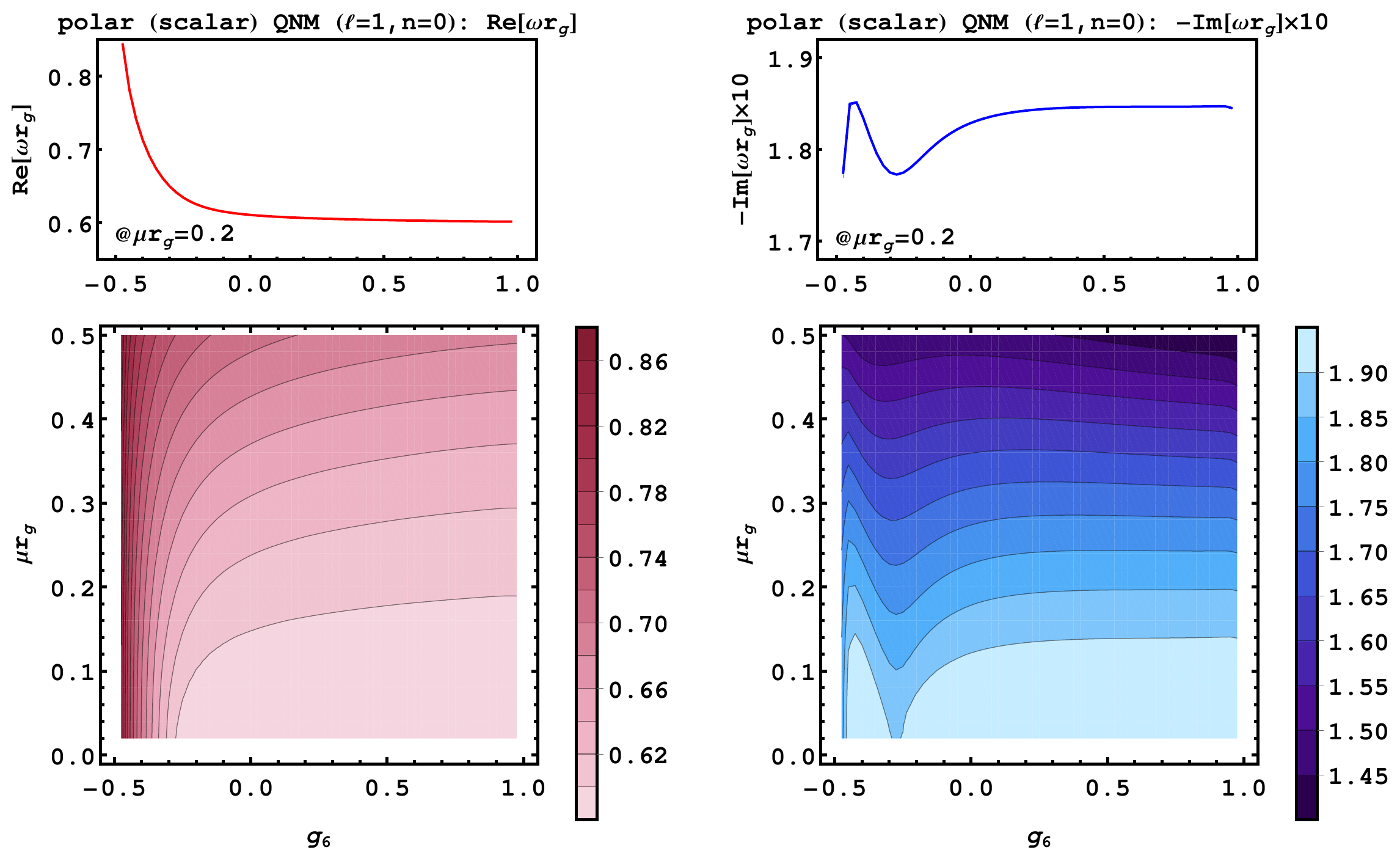}
  \caption{Polar (scalar): Real (left-hand panels) and imaginary (right-hand panels) part of the fundamental ($n=0$) first multipole ($\ell=1$) QNM in the $g_6$--$\mu$ plane (lower panels) and as a function of $g_6$ at $\mu\,r_g=0.2$.}
  \label{fig:polar_scalar_qnm_l1_n0}
\end{figure}
\begin{figure}
  \includegraphics[width=\textwidth]{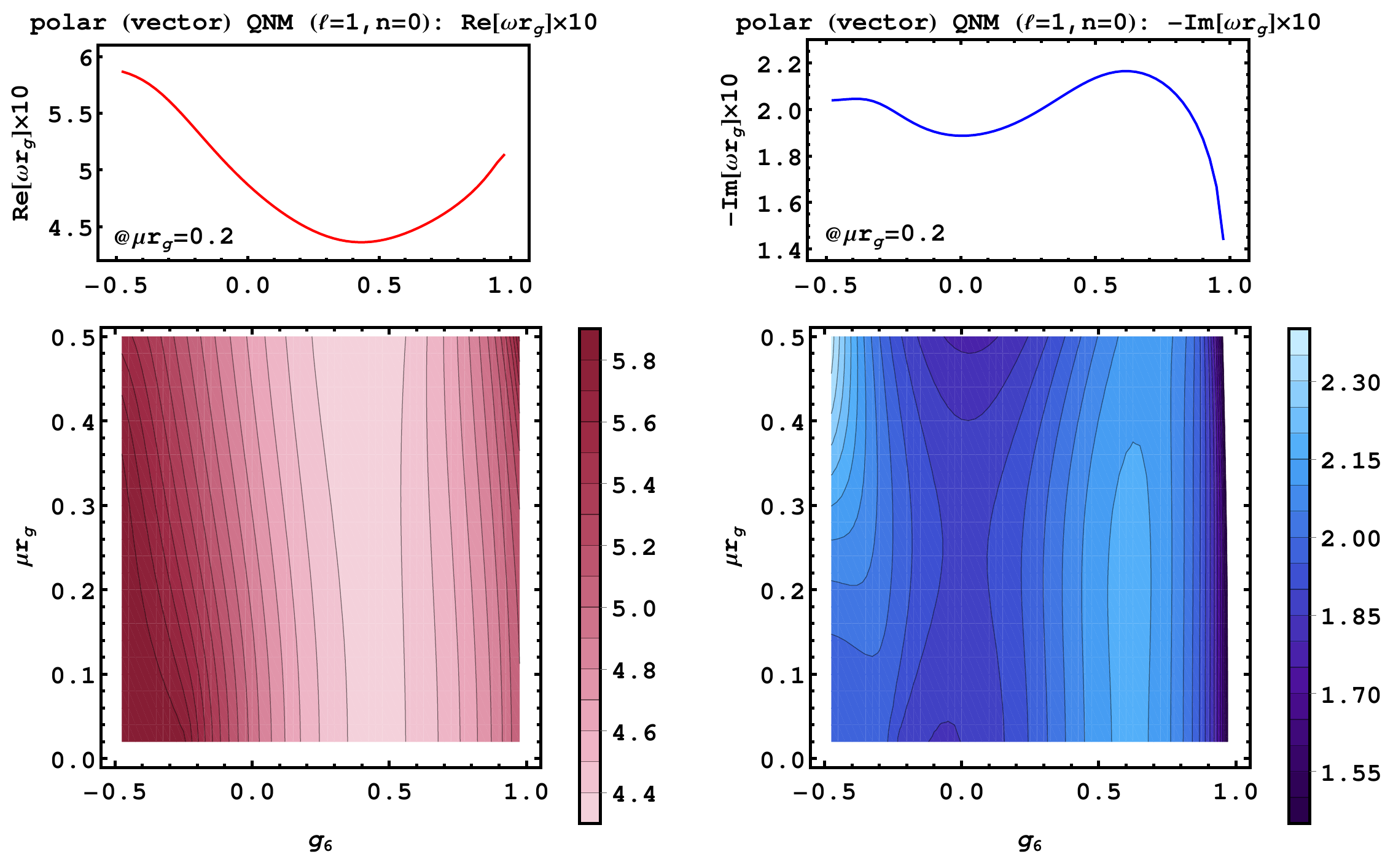}
  \caption{Polar (vector): Real (left-hand panels) and imaginary (right-hand panels) part of the fundamental ($n=0$) first multipole ($\ell=1$) QNM in the $g_6$--$\mu$ plane (lower panels) and as a function of $g_6$ at $\mu\,r_g=0.2$.}
  \label{fig:polar_vector_qnm_l1_n0}
\end{figure}
\begin{figure}
  \centering
  \includegraphics[width=0.8\textwidth]{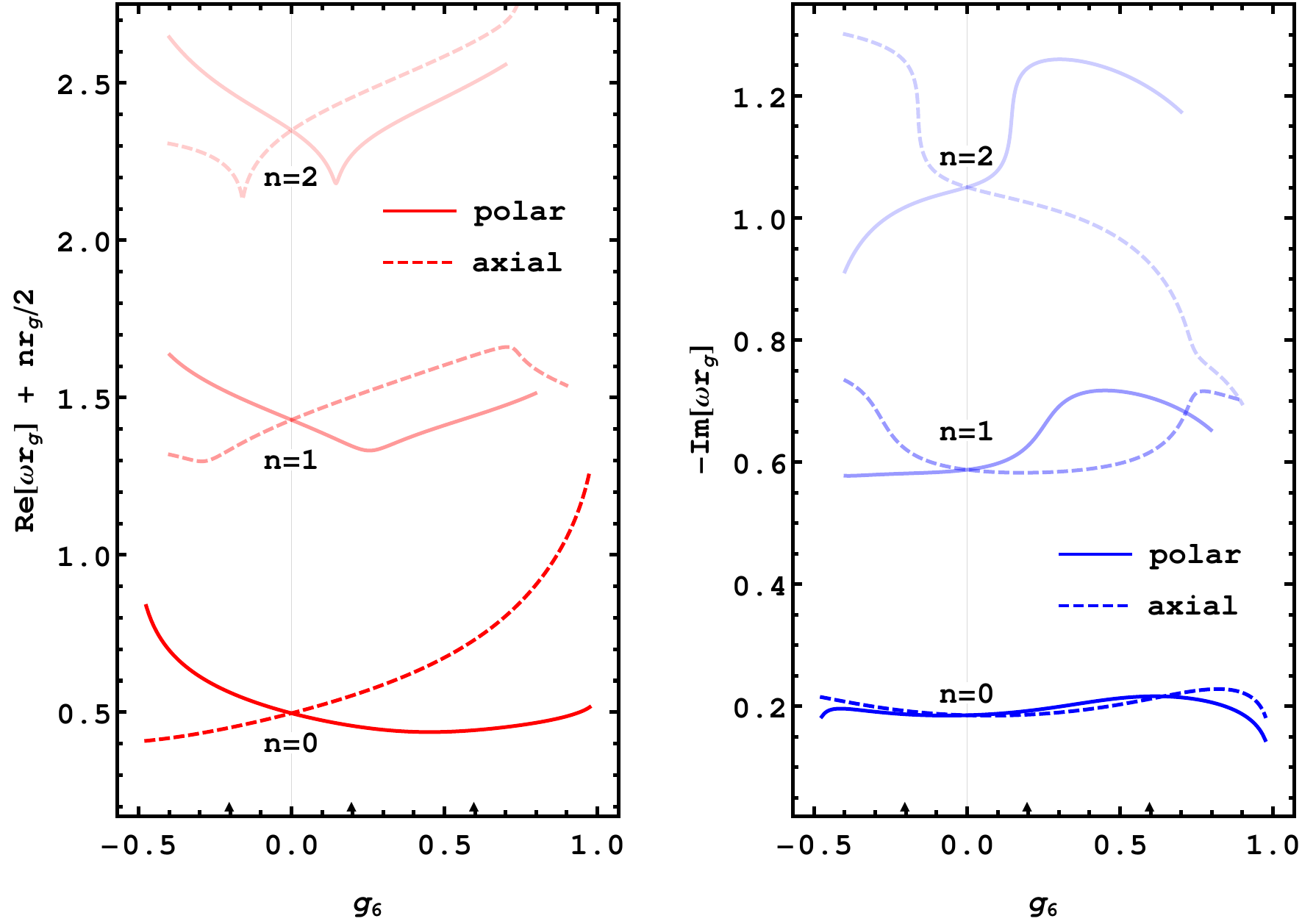}
  \caption{
  Real (left-hand panel) and imaginary (right-hand panel) part of the first multipole ($\ell=1$) QNMs in the massless case. Continuous (dashed) lines indicate polar (axial) modes. At $g_6=0$ the polar and axial mode agree (isospectrality). For any non-vanishing $g_6\neq 0$ isospectrality is broken. We show the fundamental ($n=0$) as well as first ($n=1$) and second ($n=2$) overtone in increasingly lighter shading. Where the curves end, spectral methods with $N=80$ are found to be insufficient to ensure proper convergence. Exemplary convergence plots (for the points marked with triangles on the $g_6$-axis) are presented in App.~\ref{sec:appendix}.
  }
  \label{fig:broken-isospectrality-massless}
\end{figure}

Recall that our mode equations depend on the two parameters $\mu$ and $g_6$. The standard Proca theory corresponds to $g_6=0$, whose QNM spectra on a Schwarzschild background were studied in Ref.~\cite{Rosa:2011my}. Our main aim here is the extension of the analysis to non-zero values of $g_6$ within the stability range \eqref{eq:bounds g6}, sampling also over a range of mass values $\mu$. We restrict our attention to the fundamental QNM frequency ($n=0$) and lowest multipoles $\ell=0,1$, except in the massless field case for which we present results also for the first and second overtones ($n=1,2$) of the dipole modes.

We numerically solve the mode equations using a spectral or collocation method with Chebyshev interpolation, using up to $N=80$ collocation points to ensure converged results. In essence, the method turns a differential boundary-value problem into a non-linear eigenvalue problem with finite-dimensional matrix. A brief summary of the approach is given in Appendix \ref{sec:appendix}; the reader may find a succinct but more general exposition in \cite{Dias:2015nua,Baumann:2019eav}, which also provides references to the relevant mathematical literature.

Before proceeding, a word about terminology. The polar sector contains two degrees of freedom for each $\ell\geq1$, hence two independent QNM spectra. We will refer to these modes as ``scalar'' and ``vector'', following \cite{Rosa:2011my}. The rationale behind these names is that, in the massless limit and with $g_6=0$, the polar mode equations match the form of the Regge-Wheeler (RW) equations for massless scalar and vector fields, in agreement with the Goldstone boson equivalence theorem. To see this explicitly, set $\mu=0$ and $g_6=0$, and introduce
\ba
u_2 = \frac{r}{r_g} y_2 + f \frac{r_g^2}{r^2}\frac{y_3-r y_3'}{\ell(\ell+1)} \,,\qquad  u_3 = \frac{r_g}{r}y_3 \, ,
\ea
so that the polar mode equations, Eqs.\ \eqref{eq:EoMpolar2} and \eqref{eq:EoMpolar3}, become
\ba
&&{\cal D}_{\rm RW}^{\ell,\,s=1} y_2 = 0 \,, \label{eq:eqy2} \\
&&{\cal D}_{\rm RW}^{\ell,\,s=0} y_3 - \frac{2 J}{r_g^2} y_2 = 0 \,, \label{eq:eqy3}
\ea
where
\ba
{\cal D}_{\rm RW}^{\ell,\,s}  \equiv -\frac{\pd^2}{\pd t^2} + \frac{\pd^2}{\pd r_*^2} - f\left[\frac{\ell(\ell+1)}{r^2}+\frac{(1-s^2)r_g}{r^3}\right] \,,
\ea
(with $s=0,1,2$) is the RW operator governing the dynamics of scalar, vector and tensor perturbations on the Schwarzschild spacetime. Eqs.~\eqref{eq:eqy2} and~\eqref{eq:eqy3} admit two sets of solutions: if $y_2=0$, then $y_3$ satisfies the massless scalar RW equation; if $y_2\neq0$, then $y_3$ is a pure gauge degree of freedom while $y_2$ satisfies the massless vector RW equation. These considerations can be generalized to the case with $g_6\neq0$, with the same conclusion: in the massless limit, the polar spectrum can be divided into two classes, one corresponding to a massless scalar field and another corresponding to a vector gauge field.

In Figs.~\ref{fig:mono_qnm_n0}-\ref{fig:polar_vector_qnm_l1_n0}, we present the results for the $\ell=0,1$ fundamental ($n=0$) QNMs, in case of non-vanishing mass $\mu\neq 0$. The behaviour with $0<\mu r_g<0.5$ and $-1/2<g_6<1$ is mapped out in terms of contour plots. To reveal pole-induced behaviour, we also plot the $g_6$-behaviour at fixed exemplary $\mu$. The chosen range for the vector field mass μ is motivated by the fact that one expects interesting physical effects when the Compton wavelength of the field is comparable or larger than the size of the BH, i.e. $\mu r_g\lesssim 1$. This can also be understood more mathematically from the fact that the norm of the QNM frequency can typically be estimated as $|\omega|^2 \sim V_{\rm max}$, where $V_{\rm max}$ is the height of the centrifugal potential barrier \cite{Schutz:1985km}. Now, for the QNM function to have the required wave-like behavior at spatial infinity, one also requires $|\omega| > \mu$. It follows that there will be no QNMs if $\mu^2$ is greater than the height of the centrifugal barrier, i.e. if ${\cal O}(1)$ for the lower multipoles of most physical interest.\footnote{Note that this also applies to the monopole mode, for which the role of the ``centrifugal barrier'' is played by the term $2/r^2$ in the effective potential.}

The individual results can be understood by revisiting Tab.~\ref{tab:perturbations-overview}. Each of the modes diverges at the critical values $g_6=-1/2$ and/or $g_6=+1$ iff the respective perturbation equation is affected by the corresponding pole $P_-$ ($P_+$). 
The $\ell=0$ monopole mode, cf.~Fig.~\ref{fig:mono_qnm_n0}, is affected by $P_-$ only.
The $\ell=1$ axial mode, cf.~Fig.~\ref{fig:axial_qnm_l1_n0}, is affected by $P_+$ only.
Finally, the two coupled polar multipole modes are affected by one pole each: the scalar mode, cf.~Fig.~\ref{fig:polar_scalar_qnm_l1_n0}, by $P_-$; the vector mode, cf.~Fig.~\ref{fig:polar_vector_qnm_l1_n0}, by $P_+$.
While not presenting respective results, we expect the same pole-induced behaviour to persist for all higher $\ell>1$ modes.

In case of vanishing mass $\mu=0$, the perturbations reduce to one axial and one polar mode only, cf.~Sec.~\ref{sec:massless}. For minimal coupling to the background metric, i.e., for $g_6=0$, the two QNM spectra are known to be isospectral, i.e., the axial and polar spectrum agree. Indeed, the respective perturbation equations \eqref{eq:EoMpolarmassless} and \eqref{eq:EoMaxial} (with $\mu=0$) become redundant for $g_6=0$, i.e., for $P_\pm=1$. For any $g_6\neq 0$, isospectrality is broken.
We verify this explicitly by presenting the $n=0,1,2$ massless modes in Fig.~\ref{fig:broken-isospectrality-massless}.

As in the massive case, the observed behaviour close to $g_6=-1/2$ and/or $g_6=1$ is determined by the poles in the respective perturbations equations. The axial massless mode, cf.~eq.~\eqref{eq:EoMaxial}, is affected by $P_+$ only. The polar massless mode, cf.~eq.~\eqref{eq:EoMpolarmassless}, is affected by both poles. The onset of this pole-induced behaviour can be explicitly seen for the $n=0$ mode. For the $n>1$ modes it becomes numerically challenging to resolve.

In fact, numerical convergence of the spectral methods worsens considerably with growing $n$. This can intuitively be understood as follows. The number of oscillations in $r$ increases with $n$. The modes thus become more and more challenging to resolve with spectral methods based on a fixed number of collation points. The $n>0$ results presented in Fig.~\ref{fig:broken-isospectrality-massless}, thus present the most challenging numerics of this work. Hence, we explicitly present convergence plots for exemplary points in App.~\ref{sec:appendix}.

The behavior of the QNM spectrum for small values of $g_6$ is worth remarking. As one can glean from Fig.~\ref{fig:broken-isospectrality-massless} (although we have also verified it from the numerical data), for each $n$ the polar and axial QNM frequencies display a symmetry in their $g_6$-dependence at linear order, exhibiting the same slope (within numerical precision) but with opposite sign. This feature may hint at the existence of an electromagnetic duality for small but non-zero values of $G_6$.\footnote{We thank Luca Santoni for bringing this point to our attention.} We will encounter a similar phenomenon when we consider the electromagnetic susceptibilities in Sec.\ \ref{sec:susceptibilities}.

Finally, we comment on the observed crossing of the axial $n=1$ and $n=2$ imaginary parts close to $g_6=1$. We are not aware of other examples of such a crossing of imaginary parts. While we find no indication for convergence issues in the applied spectral methods, a confirmation of this result by independent numerical techniques would be welcome. 


\section{Quasi-bound states} \label{sec:bound states}

Quasi-bound state solutions to the mode equations are defined by boundary conditions corresponding to an ingoing wave at the event horizon and a vanishing amplitude at spatial infinity. The latter requirement selects $c_{\rm out}=0$ in \eqref{eq:asymptotic sol}, oppositely to the case of QNMs. Importantly, the bound state behavior $u\sim e^{-\sqrt{\mu^2-\omega^2}\,r_{*}}$ holds regardless of the sign of ${\rm Im}\,\omega$.

The last remark is apposite given our interest in establishing whether the theory admits unstable solutions with ${\rm Im}\,\omega>0$, even if the coupling $g_6$ lies in the range \eqref{eq:bounds g6} in which {\it localized} perturbations are stable. The latter condition is necessary for consistency, as it has been shown that localized modes must be either stable or else suffer from ghost- or gradient-type instabilities, while tachyon-type unstable solutions cannot occur \cite{Garcia-Saenz:2021uyv}. The caveat to this statement is that tachyonic solutions cannot be fully diagnosed in the localized approximation, which is by definition oblivious to modes of physical size comparable or larger than the length scales of the background. In other words, we would like to assess if {\it global} solutions could undergo instabilities in the regime where the theory is free from pathologies related to ghosts and negative-gradient modes.

Here we provide strong evidence that tachyonic quasi-bound state solutions for a massive vector field cannot occur on a Schwarzschild BH background. Our first argument in support of this claim is given by the numerical results, presented in Sec.\ \ref{sec:bs numerical}, for the fundamental ($n=0$) bound state solution for each of the lowest multipole modes of the field ($\ell=0,1$), sampling over a range of values of the parameter $g_6$ and a few values of the mass $\mu$. While this certainly does not constitute a full proof, one naturally expects tachyon modes to appear for the lowest values of $n$ and $\ell$ if they exist at all.\footnote{For instance, in the case of a massive spin-2 field on a BH background it is the $n=0$, $\ell=0$ mode, and only this mode, which is unstable for a certain range of parameters \cite{Brito:2013wya,Rosen:2020crj}.} Indeed, tachyonic instabilities should, by definition, eventually disappear as the typical radial and angular wavelengths of the solutions (characterized respectively by $n$ and $\ell$) become short enough.

A more critical loophole in this numerics-based argument is our inability to access values of $g_6$ arbitrarily close to the stability bounds \eqref{eq:bounds g6}, as our numerical routine becomes increasingly less efficient as we approach those values. This is important in view of the expectation (suggested also by the numerical results) that quasi-bound state frequencies will differ the most from their values in standard Proca theory precisely near the critical $g_6$ points. Fortunately we can patch this issue by means of an analytical proof which shows that the imaginary part of the frequency cannot be positive. This argument is also not a complete one, however, first because it does not apply to the polar modes with $\ell\geq1$, and second because it assumes that $g_6$ lies sufficiently close to either of the critical points. These caveats notwithstanding, the argument is otherwise general, valid for any mass $\mu$ (with $\mu^2>0$) and for all multipoles $\ell,m$. We describe the argument and its application to the monopole and axial modes in Sec.\ \ref{sec:bs analytical}.

\subsection{Numerical results} \label{sec:bs numerical}

\begin{figure}
  \centering
  \includegraphics[width=0.9\textwidth]{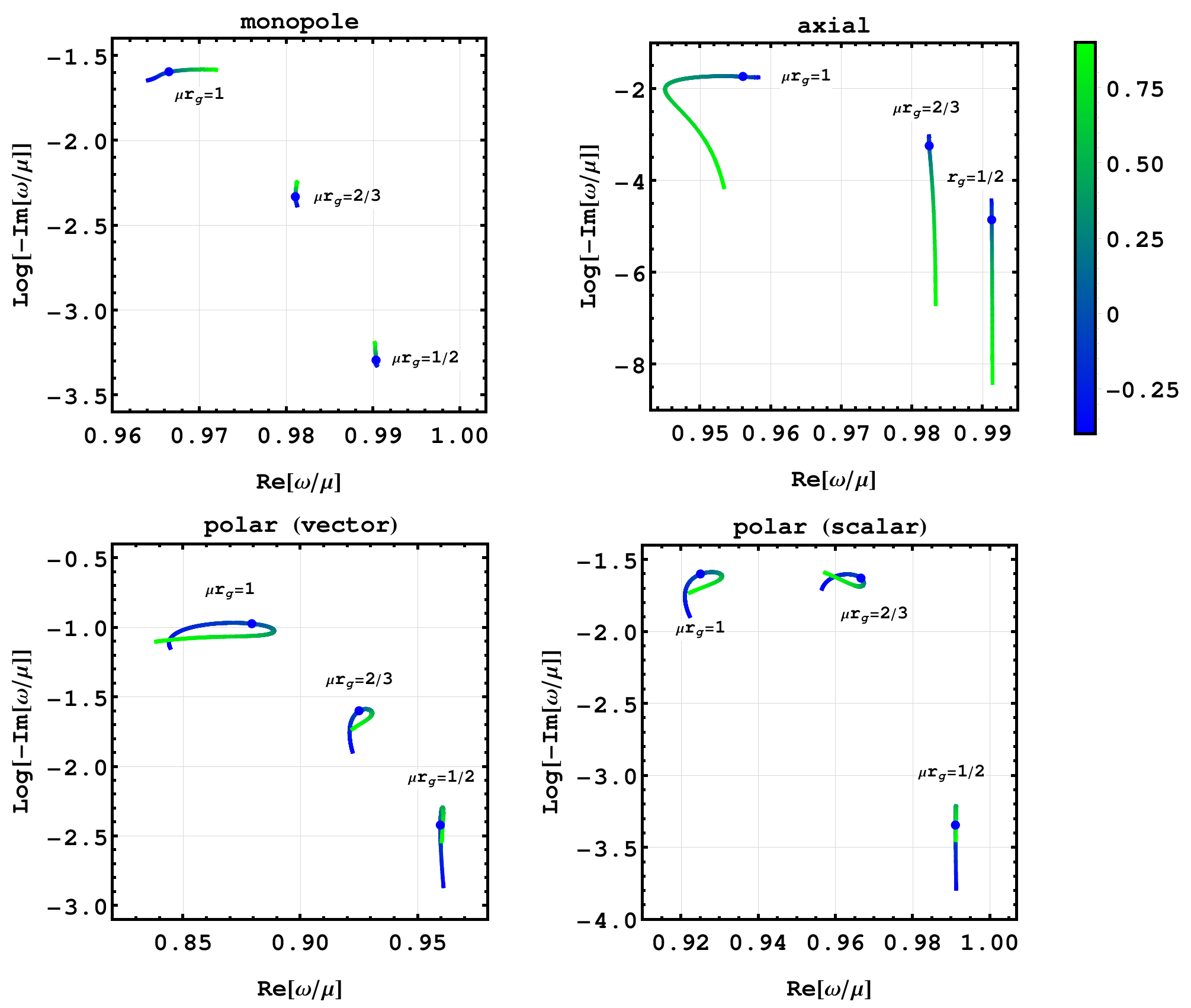}
  \caption{
  Behaviour of the fundamental $\ell=0,1$ bound states with changing $\mu$ and $g_6$. For each mode, we present the parametric curves mapped out by $-0.4<g_6<0.9$ (cf.~color legend on the right) for three different values of $\mu\times r_g=1,\,2/3,\,1/2$. The dots indicate the respective value for $g_6=0$.
  }
  \label{fig:boundStates_complex-plane}
\end{figure}

As for the case of quasi-normal modes, cf.~Sec.~\ref{sec:qnm} and App.~\ref{sec:appendix}, we numerically solve the quasi-bound state mode equations via spectral methods with Chebyshev interpolation. We restrict the analysis to the fundamental monopole ($\ell=0$) and lowest multipole ($\ell=1$) modes. We make sure that all of the following results are converged to at least $5\%$ accuracy. (For most parameter values the accuracy is much higher, cf.~App.~\ref{sec:appendix} for exemplary convergence plots in the QNM case.)

In Fig.~\ref{fig:boundStates_complex-plane}, we summarize the behavior of the fundamental $\ell=0,1$ quasi-bound states in the complex-frequency plane.\footnote{With some abuse of terminology, we refer to the polar $\ell=1$ modes as ``scalar'' and ``vector'' as we did for QNMs, although in reality quasi-bound states cease to exist in the massless limit and therefore the Goldstone boson equivalence limit is not meaningful.} To do so, we show the quasi-bound states for $-0.4<g_6<0.9$ and representative $\mu\times r_g=1,\,2/3,\,1/2$. With $\mu\rightarrow 0$, these all converge to $\text{Re}[\omega/\mu]\rightarrow 1$ and $0>\text{Im}[\omega/\mu]\rightarrow 0$. This holds for any constant $g_6$, at least in the investigated range. 
For the axial mode, cf.~upper-right panel in Fig.~\ref{fig:boundStates_complex-plane}, we find $\text{Im}[\omega/\mu]\stackrel{g_6\rightarrow 1}{\longrightarrow} 0$, for all investigated values of $\mu$. We find no indications for the onset of such scaling for the other modes, at least within the investigated range.

\begin{figure}
  \centering
  \includegraphics[width=0.85\textwidth]{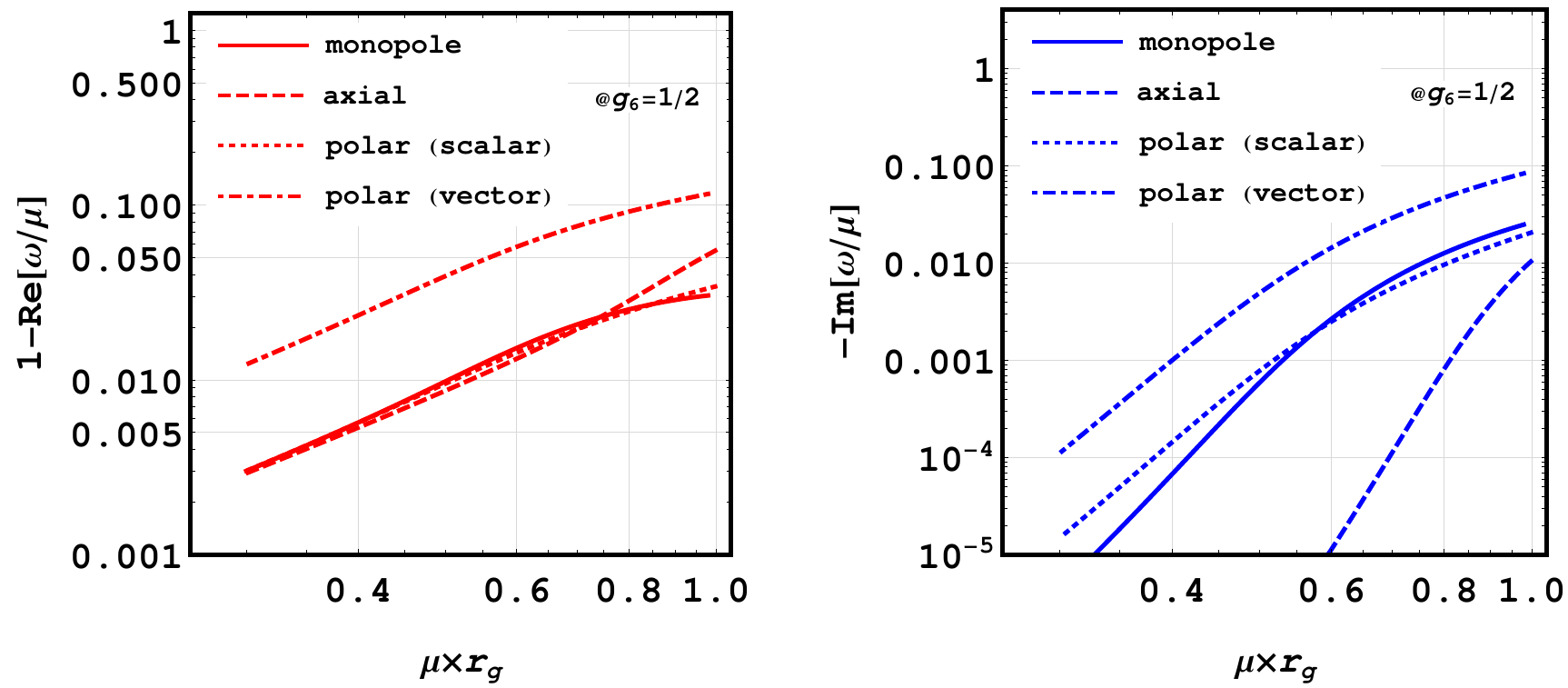}
  \caption{
  Power-law behaviour of the fundamental $\ell=0,1$ bound states with $\mu\rightarrow 0$ at an exemplary value of $g_6=1/2$.
  }
  \label{fig:boundStates_powerlaw}
\end{figure}

In Fig.~\ref{fig:boundStates_powerlaw}, we exemplify the power-law behaviour that all modes exhibit as $\mu\rightarrow 0$. Here, we choose to present results at a representative value of $g_6=1/2$ only. The behaviour closely resembles the one previously found for $g_6=0$ \cite{Rosa:2011my}.

To summarize, we find no indication for the presence of an unstable mode (i.e., one with $\text{Im}[\omega]>0$). Whenever modes scale towards $\text{Im}[\omega]=0$, we have identified the respective power-law scaling. We view this as strong numerical evidence for the absence of unstable quasibound states.

\subsection{Integral formula} \label{sec:bs analytical}

Next we turn to the analytical proof of the fact that ${\rm Im}\,\omega<0$ in our set-up. The method is essentially the one put forth in Ref.~\cite{Horowitz:1999jd} in the context of asymptotically anti-de Sitter BHs (see also Ref.~\cite{Cardoso:2001bb} for further applications). The interesting observation is that the argument also applies to bound state perturbations of asymptotically flat BHs, albeit with some differences.

We consider eq.\ \eqref{eq:generic mode eq} in the case of a single ODE, so that $\mathcal{V}(u,r)\equiv V(r)u$. We introduce the redefined mode function $v\equiv e^{i\omega r_*}u$. The boundary conditions imply that $v$ approaches a constant, $v_+$, at the horizon and that it decays exponentially at spatial infinity. The mode equation for $v$ takes the form
\beq
(fv')'-2i\omega v'-Vv=0 \,.
\eeq
We multiply through by $v^{*}$ and integrate,
\beq
\int_{r_g}^{\infty}dr\left[v^{*}(fv')'-2i\omega v^{*}v'-V|v|^2\right]=0 \,.
\eeq
Note that each term in this integral gives a finite result thanks to the exponential decay of $v$ (while $V$ is non-singular by assumption). The first term can be integrated by parts, noting that the boundary term vanishes,
\beq \label{eq:int argument result1}
\int_{r_g}^{\infty}dr\left[f|v'|^2+2i\omega v^{*}v'+V|v|^2\right]=0 \,.
\eeq
Taking the difference of this equation with its complex conjugate we get
\beq
\int_{r_g}^{\infty}dr\,v^{*}v'=-\frac{\omega^{*}}{2i\,{\rm Im}\,\omega}\,|v|^2\Big|^{\infty}_{r_g}= \frac{\omega^{*}|v_{+}|^2}{2i\,{\rm Im}\,\omega} \,,
\eeq
which can be plugged back in \eqref{eq:int argument result1} to produce
\beq \label{eq:int argument result2}
\int_{r_g}^{\infty}dr\left[f|v'|^2+V|v|^2\right]=-\frac{|\omega|^2|v_{+}|^2}{{\rm Im}\,\omega} \,.
\eeq
We see that if the potential function $V$ were positive definite, then we would immediately infer that ${\rm Im}\,\omega<0$ and conclude the proof. However $V$ is not positive definite in the equations within our set-up. Nevertheless, we can prove that, for each mode, its contribution to the integral is indeed non-negative whenever $g_6$ is sufficiently close to the critical points, i.e.,\ for values such that the poles $P_{\pm}$ coincide with the event horizon.

The effective potential of the monopole mode is given by
\beq
V_M(r)=\frac{\mu^2}{P_-}+\frac{2}{r^2}-\frac{3r_g}{r^3} \,,
\eeq
and it is easy to see that $V_M$ is not positive definite for all values of $\mu^2$ and $g_6$. However, as we are interested in the case when $g_6$ lies near the bound $g_6 =-1/2$, we define $\epsilon\equiv g_6+1/2$ and isolate the leading-order contribution to the integral \eqref{eq:int argument result2} in an expansion in small $\epsilon$, i.e.,
\beq
\int_{r_g}^{\infty}dr\, V_M|v|^2\simeq \mu^2|v_{+}|^2\int_{r_g}^{\infty}\frac{dr}{(r/r_g)^3-(1-2\epsilon)}\simeq \frac{\mu^2|v_+|^2r_g}{3}\,\log\frac{1}{\epsilon} \,.
\eeq
This integral is manifestly positive. 

Similarly, for the axial modes the effective potential reads
\beq
V_{-}(r)=\frac{\mu^2}{P_{+}}+\frac{\ell(\ell+1)}{r^2}\,\frac{P_-}{P_+}+\frac{9}{4r^2}\left(1-\frac{1}{P_+}\right)\left(\frac{f}{P_+}+\frac{5}{3}-\frac{7r_g}{3r}\right) \,,
\eeq
which is also not positive definite for all $\mu$, $\ell$ and $g_6$. The relevant pole is now $g_6=1$, so we let $\epsilon\equiv 1-g_6$ and evaluate the integral to leading order in the limit of small $\epsilon$,
\beq
\int_{r_g}^{\infty}dr\, V_{-}|v|^2\simeq  \frac{|v_+|^2}{r_g}\left(\frac{\mu^2r_g^2}{3}+\ell(\ell+1)+\frac{1}{4}\right)\log\frac{1}{\epsilon} \,,
\eeq
and the result is likewise manifestly positive.

This establishes that ${\rm Im}\,\omega<0$ for quasi-bound state perturbations corresponding to monopole and axial modes. For the polar modes with $\ell\geq1$ the argument does not readily apply since in this case one has to deal with a system of coupled equations and with additional terms proportional to derivatives of the mode functions, cf.\ eq.\ \eqref{eq:EoMpolar3}. Even though an analogue of eq.\ \eqref{eq:int argument result2} can be straightforwardly derived, we have been unable to find a bound for the integral of the resulting effective potential. Nevertheless, we see no reason why polar perturbations should behave qualitatively different from the rest of the spectrum, and the numerical results certainly seem to confirm this. Moreover, as we remarked previously, the expectation is that unstable modes, if they exist, should manifest themselves at the lower end of the multipole ladder. Given our proof of the stability of monopole fluctuations, we take these combined results as strong evidence for the absence of instabilities in the whole quasi-bound state spectrum and the whole range of allowed values of the non-minimal coupling $g_6$.


\section{Electromagnetic susceptibilities} \label{sec:susceptibilities}

Static response coefficients characterize the change of a system under an external time-independent field. For a gravitational field, these coefficients correspond to the tidal Love numbers, which are in principle directly measurable through gravitational wave observations, e.g.\ of binary systems. For a $U(1)$ gauge field the response coefficients are the electric and magnetic susceptibilities defining the polarizability of the object (in analogy with electromagnetism, although the field of course need not be the Standard Model photon).

As mentioned in the introduction, a remarkable property of four-dimensional BHs in GR is that they do not polarize under the effects of a Maxwell-type field. Yet the expectation is that this attribute will be broken in more general set-ups, in particular if the external $U(1)$ field contains additional interactions, either with itself or with the spacetime metric. Within our set-up of a Schwarzschild BH and in linear response theory, we have seen that it is only the Horndeski non-minimal coupling operator, eq.\ \eqref{eq:g6 coupling}, which can contribute to beyond-GR effects without introducing additional degrees of freedom. The question is then whether the electromagnetic susceptibilities are indeed non-vanishing when the coupling $G_6$ is non-zero. Here we confirm that they are non-vanishing, focusing for simplicity in the case of dipolar perturbations.

\subsection{Boundary expansion}

We consider the mode equations in the gauge invariant case, i.e.\ eq.\ \eqref{eq:EoMpolarmassless} for the polar or ``electric'' field and eq.\ \eqref{eq:EoMaxial} (with $\mu=0$) for the axial or ``magnetic'' field, setting $\omega=0$ as we are interested in the static limit.

For each equation, only one linear combination of the two independent solutions is regular at the event horizon. Demanding regularity thus fixes one integration constant, while the other remains arbitrary, simply setting the overall amplitude of the mode function. Then, modulo this overall constant, the solution at spatial infinity is fully determined, and is in general given by a sum of two modes, one which grows and one which decays with the radius $r$, i.e.,
\beq \label{eq:suscept bdry expansion}
u(r)=c_{\rm ext}\left(\frac{r}{r_g}\right)^{\ell+1}\big(1+\mathcal{O}(r_g/r)\big)+c_{\rm resp}\left(\frac{r}{r_g}\right)^{-\ell}\big(1+\mathcal{O}(r_g/r)\big) \,.
\eeq
The leading coefficient of the growing mode, $c_{\rm ext}$, is interpreted as the strength of the applied field, while that of the decaying mode, $c_{\rm resp}$, gives the corresponding response of the system. Their ratio,
\beq \label{eq:suscept def}
k\equiv \frac{c_{\rm resp}}{c_{\rm ext}} \,,
\eeq
defines the linear susceptibility of the system for the given external field.

There are two remarks to keep in mind about the structure of the boundary expansion in eq.\ \eqref{eq:suscept bdry expansion}, both related to the fact that the expansion is a Frobenius series. The first is that the series multiplying $r^{\ell+1}$ in the growing mode may in general contain logarithmic terms. However, in four dimensions these are always subleading and do not affect the definition in \eqref{eq:suscept def}.\footnote{This is not necessarily the case in spacetime dimension other than four, where the logarithmic terms may induce a renormalization group running of the response coefficients \cite{Kol:2011vg,Hui:2020xxx}.} The second observation is that, because $\ell$ is an integer, the split between the growing and decaying modes is potentially ambiguous as they contain the same powers of $r$ after some order \cite{Fang:2005qq,Pani:2015hfa}. Various ways to deal with this issue have been proposed in the literature \cite{Binnington:2009bb,Kol:2011vg,Chakrabarti:2013lua,Poisson:2020vap}, although in our case it will suffice to simply define the growing mode such that it does not contain the power $r^{-\ell}$ (which is not to say that the series terminates, since all subsequent powers may a priori be present). This prescription makes the susceptibility $k$ unambiguous and is physically justified by the fact that $k$ so defined is an observable enjoying the property we seek: it vanishes in the absence of non-minimal coupling but is otherwise non-zero, as we now show.

For illustration, let us focus on the dipole modes ($\ell=1$). Interestingly, we find that there is no logarithmic term in this case. However, unlike in the ordinary Maxwell set-up, the series for the growing mode does not terminate. Explicitly, for the first few terms we find
\beq\bal
u_E(r)&=c_{\rm ext}\left(\frac{r}{r_g}\right)^2+c_{\rm resp}\left(\frac{r}{r_g}\right)^{-1}+\frac{3(2c_{\rm resp}-5g_6c_{\rm ext})}{8}\left(\frac{r}{r_g}\right)^{-2} \\
&\quad +\frac{3(2c_{\rm resp}-5g_6c_{\rm ext})}{10}\left(\frac{r}{r_g}\right)^{-3}+\frac{4c_{\rm resp}-g_6(10-11g_6)c_{\rm ext}}{8}\left(\frac{r}{r_g}\right)^{-4}+\mathcal{O}(r/r_g)^{-5} \,,
\eal\eeq
\beq\bal
u_M(r)&=c_{\rm ext}\left(\frac{r}{r_g}\right)^2+c_{\rm resp}\left(\frac{r}{r_g}\right)^{-1}+\frac{3(2c_{\rm resp}+5g_6c_{\rm ext})}{8}\left(\frac{r}{r_g}\right)^{-2} \\
&\quad +\frac{3(2c_{\rm resp}+5g_6c_{\rm ext})}{10}\left(\frac{r}{r_g}\right)^{-3}+\frac{4c_{\rm resp}+g_6(10-g_6)c_{\rm ext}}{8}\left(\frac{r}{r_g}\right)^{-4}+\mathcal{O}(r/r_g)^{-5} \,,
\eal\eeq
respectively for the electric and magnetic components.

Although the mode equations do not seem to admit an exact solution, they may straightforwardly be solved iteratively by expanding in powers of the coupling $g_6$.\footnote{We recall that $g_6$ enters in the equations through the combinations $P_{\pm}=1-r_{\pm}^3/r^3$ (cf.\ \eqref{eq:poles}), where $r_{\pm}<r_g<r$ and $r_{\pm}^3\propto g_6$. Therefore, for any $r$ in the physical domain, the mode equations are indeed analytic at $g_6=0$ and the expansion in Taylor series is justified.} After selecting the regular solution in each case, as explained above, we are then able to infer $k$. We find, up to order $\mathcal{O}(g_6^4)$,
\beq \label{eq:suscept conjecture}
k_E=\frac{5}{2}\,g_6+\mathcal{O}(g_6^5) \,,\qquad k_M=-\frac{5}{2}\,g_6+g_6^2-\frac{6}{7}\,g_6^3+\frac{209}{280}\,g_6^4+\mathcal{O}(g_6^5) \,,
\eeq
respectively for the electric and magnetic susceptibilities.

The expressions in \eqref{eq:suscept conjecture} confirm the vanishing of the susceptibility in standard Maxwell theory, i.e.\ with $g_6=0$. For small but non-zero $g_6$ we find instead the expected dependence $k=\mathcal{O}(g_6)$. We observe that the electric and magnetic susceptibilities are equal, up to a sign, at linear order in $g_6$. We recall that the same phenomenon was observed for the QNM spectrum in the massless (i.e., gauge-invariant) case, cf.~Fig.~\ref{fig:broken-isospectrality-massless}. Also remarkable is that the electric susceptibility does not appear to receive non-linear corrections. These results are intriguing and clearly beg for a deeper physical understanding. We hope to come back to this question in future work.

\subsection{Numerical results}

Having understood analytically the polarizability properties of a BH in the approximation of small $g_6$, we now turn to the exact results derived numerically using a shooting method. We have computed the solutions of the mode equations in the vicinity of the BH horizon by expanding in powers of $(r-r_g)$, up to order four. We then evaluate the function at some small $(r-r_g)$, which is used as initial condition to integrate numerically up to some large radius. The result is then matched to the boundary series discussed in the previous subsection, which we expand up to order $r^{-8}$, so as to obtain $c_{\rm ext}$ and $c_{\rm resp}$, and hence the susceptibility $k$. We have checked that the results are robust against changes in the initial and matching radii as well as in the order at which we terminate the series ansatze.

\begin{figure}
	\centering
	\includegraphics[width=1\textwidth]{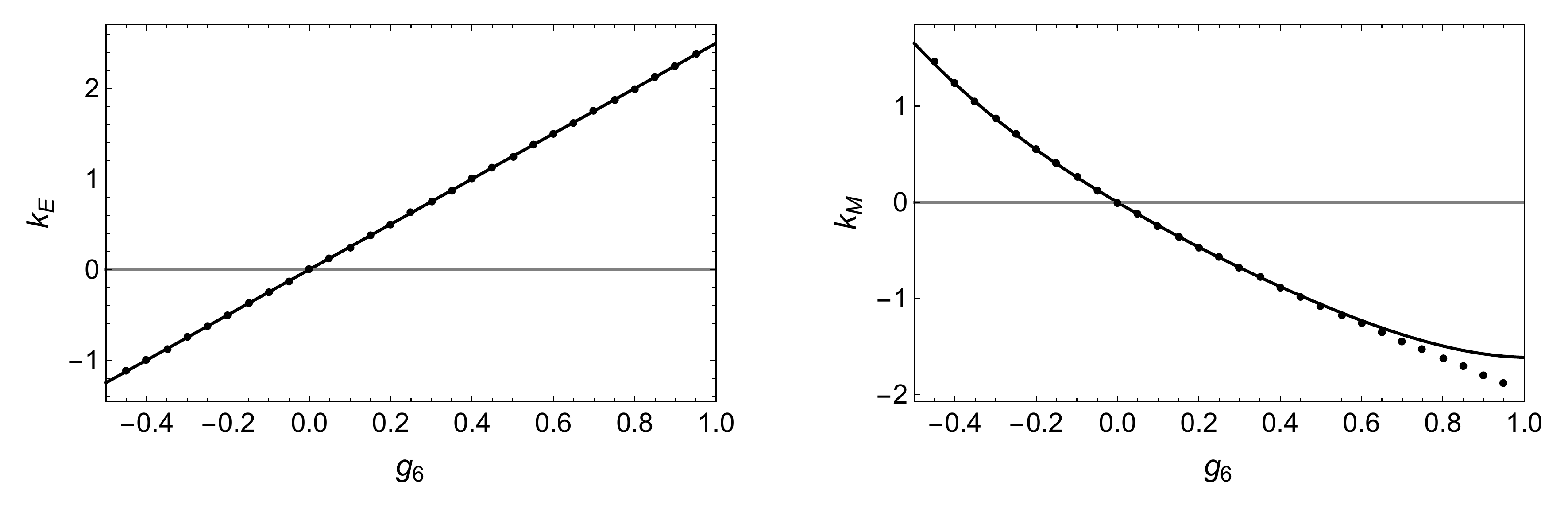}
	\caption{Dipolar electric (left) and magnetic (right) susceptibilities as functions of the non-minimal coupling $g_6$. Dots are the numerical results, solid lines correspond to the analytical approximations in eq.\ \eqref{eq:suscept conjecture}.}
	\label{fig:suscept}
\end{figure}

The results for the electric and magnetic susceptibilities are shown in Fig.\ \ref{fig:suscept}, plotted as functions of the coupling $g_6$ within the stability range eq.~\eqref{eq:bounds g6}. The plots also show the comparison with the approximate analytical behaviors in eq.~\eqref{eq:suscept conjecture}, which are indeed in perfect agreement with the numerical results. For the electric susceptibility we confirm the interesting outcome that the linear truncation in eq.~\eqref{eq:suscept conjecture} appears to be {\it exact}, within our numerical precision, even as we get very close to the critical values of $g_6$ (we can reliably compute $k$ for $|g_6-g_6^{\rm crit}|\gtrsim10^{-3}$). In contrast, the magnetic susceptibility shows a clear departure from the polynomial approximation for sizable values of $g_6$, as one would generically expect\footnote{From the truncated Taylor series for $k_M$ in eq.~\eqref{eq:suscept conjecture} one may also construct a Pad\'e approximant in order to get a better fit of the numerical results. We thank Hector Silva for pointing this out to us.}.
%


\section{Discussion} \label{sec:discussion}

The aim of this paper was to initiate the study of global solutions for massive vector fields non-minimally coupled to gravity in the linear approximation about GR backgrounds. We focused on the simplest but physically important case of a Schwarzschild BH background, and restricted our attention to a single non-minimal coupling operator, namely the Horndeski term given in eq.\ \eqref{eq:g6 coupling}. In spite of the simplicity of the model under consideration, we showed that the set-up is in fact unique, in the sense that any vector-tensor Lagrangian must reduce to \eqref{eq:full lagrangian} upon linearization of the vector field about the vacuum $\langle A_{\mu}\rangle=0$, assuming the theory describes $3+2$ dynamical degrees of freedom.

Our principal result is the outcome of the numerical calculation of the fundamental QNM frequency for the lowest multipole modes of the vector field, i.e.\ monopole (Fig.\ \ref{fig:mono_qnm_n0}), axial dipole (Fig.\ \ref{fig:axial_qnm_l1_n0}), and polar scalar and vector dipoles (Figs.\ \ref{fig:polar_scalar_qnm_l1_n0}, \ref{fig:polar_vector_qnm_l1_n0}). We explored a physically motivated range of values for the Proca mass $\mu$, as well as the full range for the (normalized) non-minimal coupling parameter $g_6$ allowed by stability. However, our results exclude values very close to the bounds, eq.\ \eqref{eq:bounds g6}, where our numerics become unreliable. It would be desirable to gain a better grasp on the behavior of QNMs when $g_6$ is at or arbitrarily close the critical values. This is not merely an academic question, since we recall that $g_6\equiv G_6/r_g^2$ depends on the BH mass, so for any given non-zero coupling $G_6$ there will be a BH mass value such that either of the bounds is saturated. Of course, whether such a BH mass is physical, and whether the theory at that scale still makes sense, is a different question.

In the case where the vector field is massless the set-up simplifies considerably thanks to the $U(1)$ gauge symmetry of the theory, and one is left with a single mode (for each $\ell,m$) in each of the polar and axial sectors, i.e.\ the analogs of the electric and magnetic fields, allowing us also to compute the first two overtone QNM frequencies ($n=1,2$) as functions of $g_6$. One interesting, although perhaps not unexpected, conclusion is that the isospectrality between polar and axial QNMs is broken by the non-minimal coupling, cf.\ Fig.\ \ref{fig:broken-isospectrality-massless}.

Another set of valuable observables in the gauge-invariant setting is given by the electromagnetic susceptibilities, corresponding to the linear response of the BH to a static external field. While for a minimally coupled $U(1)$ field BHs in GR do not polarize, as recalled in the introduction, our results demonstrate that this property ceases to hold in the presence of the Horndeski non-minimal coupling that we studied. We have shown this here explicitly for the dipole modes, cf.\ Fig.\ \ref{fig:suscept}, for which we also provided some analytical understanding of the dependence of the susceptibility coefficients at linear order in the parameter $g_6$, cf.\ eq.\ \eqref{eq:suscept conjecture}. We plan to undertake a more general analysis in a dedicated work.

The question on the stability of astrophysically relevant GR backgrounds under fluctuations of generalized vector fields motivated us also to study quasi-bound state solutions within our set-up. Unlike QNMs in asymptotically flat spacetimes, quasi-bound state frequencies may in principle develop a positive imaginary part, signaling a tachyon-type instability. Whether this indeed can occur for vector fields is an important issue because a tachyonic destabilization is a possible mechanism to generate compact astrophysical objects with vector hair starting from a hairless initial state---a phenomenon known as {\it vectorization} \cite{Ramazanoglu:2017xbl,Ramazanoglu:2018tig,Annulli:2019fzq,Minamitsuji:2020pak}. The no-go result of \cite{Garcia-Saenz:2021uyv}, together with the more general analyses in \cite{Silva:2021jya,Demirboga:2021nrc}, cast doubt on vectorization as a viable mechanism, as they showed that localized perturbations must be either stable or else grow through wrong-sign kinetic or gradient operators. Our present results further supplement this claim by demonstrating that global bound-state solutions on a Schwarzschild BH background likewise do not exhibit tachyonic growth. We warn the reader that our argument does have some potential loopholes, as we explained at length in Sec.\ \ref{sec:bs analytical}, although they are not expected to be critical. As an incidental outcome of our analysis, we also showed how an integral formula for the imaginary part of the quasi-bound state frequency due to Horowitz and Hubeny may be applied to asymptotically flat spacetimes. We hope to revisit this problem in a more general setting, e.g.\ by including matter and spin, in future investigations.

\acknowledgments

We would like to thank Luca Santoni, Hector Silva and Shuang-Yong Zhou for useful conversations and comments. The work of SGS and JZ at Imperial College London was supported by the European Union's Horizon 2020 Research Council grant 724659 MassiveCosmo ERC-2016-COG. The work of AH at Imperial College London was supported by the Royal Society International Newton Fellowship NIF{\textbackslash}R1{\textbackslash}191008. AH also acknowledges that the work leading to this publication was supported by the PRIME programme of the
German Academic Exchange Service (DAAD) with funds from the German Federal Ministry of Education and Research (BMBF). SGS thanks the Peng Huanwu Center for Fundamental Theory at USTC for generous hospitality. JZ is also supported by a scientific research starting grant No.~118900M061 from the University of Chinese Academy of Sciences.


\appendix

\section{Numerical computation of quasi-normal modes and quasi-bound states} 
\label{sec:appendix}

\subsection{Method}

In this appendix we describe the numerical method used to compute QNMs and quasi-bound states in this paper.

In Sec.~\ref{sec:decomp} we have decomposed the Proca equation into its angular and radial components. The QNMs and quasi-bound states can be found by solving the non-linear eigenvalue problem for the radial equations~\eqref{eq:EoMmono},~\eqref{eq:EoMaxial},~\eqref{eq:EoMpolar2},~\eqref{eq:EoMpolar3} and \eqref{eq:EoMpolarmassless}, supplemented with the appropriate boundary conditions discussed in Sec.~\ref{sec:bc}. For the mode equations to be amenable to our numerical routine, it is useful to factor out the wave behavior at the boundaries. This is achieved by redefining
\beq \label{eq:uansatz}
u(\omega,r)= \left(1-\frac{r_g}{r}\right)^{-i\omega r_g}r^{\pm\frac{(\mu^2-2\omega^2)r_g}{2\sqrt{\mu^2-\omega^2}}}e^{\pm\sqrt{\mu^2-\omega^2}\,r}B(r) \,,
\eeq
with the choice of $+$ sign for QNMs and $-$ sign for quasi-bound states, and where $B(r)$ is a regular function of $r$ (also implicitly of $\omega$) that tends to constant values as $r_*\to\pm\infty$. In the case of axial perturbations, as well as monopole and massless polar perturbations, the radial equation can be written as a single second order differential equation for the function $B(r)$. In order to compute the eigenfrequencies, we first approximate the differential equations with finite-dimensional matrix equations using a collocation method with Chebyshev interpolation.

We firstly introduce
\begin{align}
	\xi = \frac{r- 2\sqrt{rr_g}}{r},
\end{align}
so that the function $B(\xi)$ is defined on the finite interval $\xi \in [-1,1]$, while its equation can be written in the form
\begin{align}\label{eq:EoMB}
	\left(\frac{\rmd^2}{\rmd\xi^2} + C_1(\omega,\, \xi) \frac{\rmd}{\rmd\xi} + C_2(\omega,\, \xi)\right) B(\xi) = 0 \, .
\end{align}
Note that one can choose other mappings $\xi(r)$, the only requirement being that the singularities introduced by the non-minimal coupling terms are far enough from the domain of $\xi$ such that they do not dramatically affect the convergence. Now we expand $B(\xi)$ in terms of a set of cardinal polynomials $p_k(\xi)$,
\begin{align}
	B_N(\xi) = \sum_{k=0}^{N} B(\xi_k) p_k(\xi) \,,
\end{align}
where $p_k(\xi)$ is defined by $p_k(\xi_n) = \delta_{nk}$, and $\xi_n$ are the Chebyshev nodes,
\begin{align}
	\xi_n \equiv \cos \left(\frac{\pi(2n+1)}{2N+2}\right), \quad {\rm with} \quad n = 0,1, \ldots, N\,.
\end{align}
Since $B(\xi)$ is smooth on $\xi \in [-1,1]$, the $B_N(\xi)$ converge to $B(\xi)$ as $N$ approaches infinity. Eq.~\eqref{eq:EoMB} can thus be approximated by the algebraic system
\begin{align}\label{eq:MB}
	\sum_{k=0}^N {\cal M}_{nk}(\omega) B(\xi_k) =0 \,,
\end{align}
where
\begin{align}
	{\cal M}_{nk}(\omega) \equiv p_k''(\xi_n) + C_1(\omega, \xi_n) p_k'(\xi_n) +C_2(\omega,\xi_n)\delta_{nk} \,.
\end{align}
The derivative matrices $p_k''(\xi_n)$ and $p_k'(\xi_n)$ can be computed using the second barycentric form \cite{Berrut, Higham2004TheNS}, explicitly
\ba
p_k'(\zeta_n) &=&  \left\{
             \begin{array}{lr}
             \frac{w_k/w_n}{\zeta_n - \zeta_k} & \quad n \neq k   \\
            -\sum_{k\neq n} p_k'(\zeta_n) & \quad n = k  
             \end{array}
\right. \, , \\
p_k''(\zeta_n) &=&  \left\{ 
             \begin{array}{lr}
            2 p_k'(\xi_n)p_n'(\xi_n) -\frac{2p_k'(\xi_n)}{\xi_n -\xi_k}& \quad n \neq k   \\
            -\sum_{k\neq n} p_k''(\zeta_n) & \quad n = k  
             \end{array}
\right. \, .
\ea
With a good initial guess on the eigenfrequency, in our case e.g.\ the eigenfrequency of the standard Proca field \cite{Rosa:2011my}, one can solve Eq.~\eqref{eq:MB} for $\omega$ and the set $B(\xi_k)$.

For the massive polar modes, we instead have two coupled differential equations, say for $B_2(r)$ and $B_3(r)$, after we make the ansatz~\eqref{eq:uansatz} for $u_2$ and $u_3$. The procedure is nevertheless the same, i.e.\ we approximate the two differential equations with a set of algebraic equations of the form \eqref{eq:MB}, now with
\begin{align}
B(\xi_k) = \begin{pmatrix}
B_2(\xi_k) \\
B_3(\xi_k)
\end{pmatrix}
\end{align}
and with the matrix ${\cal M}_{nk}(\omega)$ being enlarged accordingly.

\subsection{Accuracy checks}

\begin{figure}
	\centering
	\includegraphics[width=0.485\textwidth]{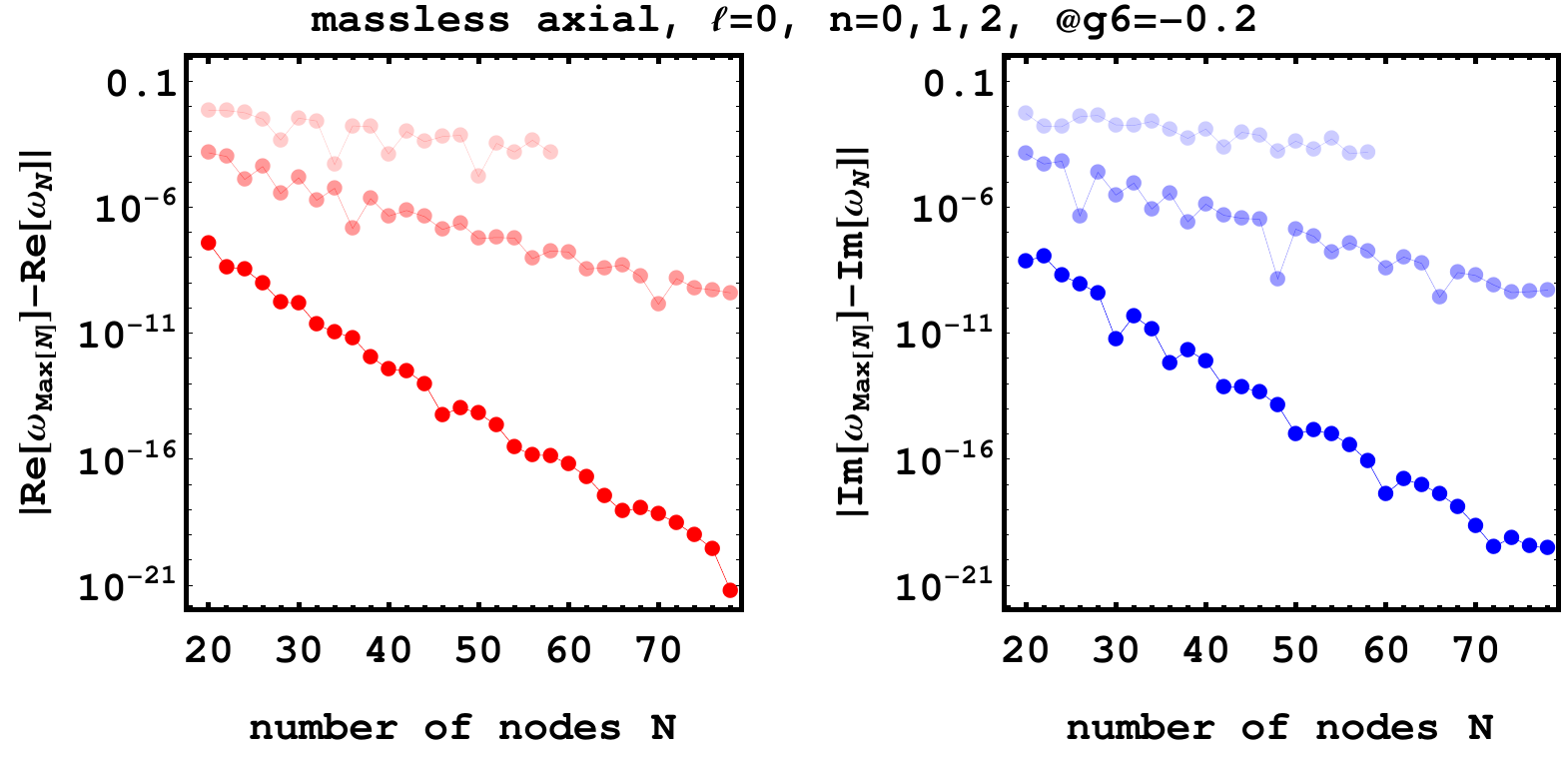}
	\hfill
	\includegraphics[width=0.485\textwidth]{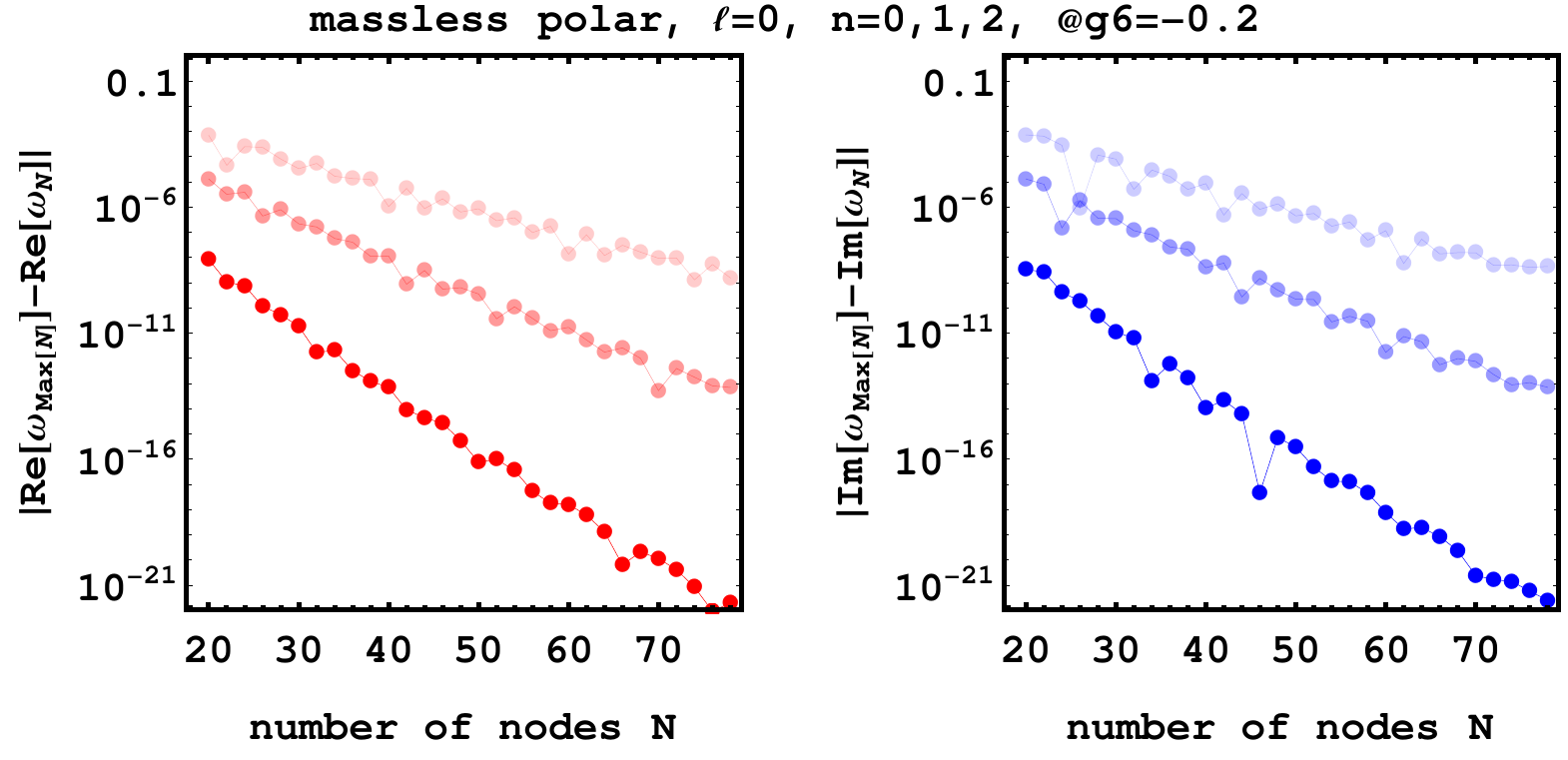}
	\\
	\includegraphics[width=0.485\textwidth]{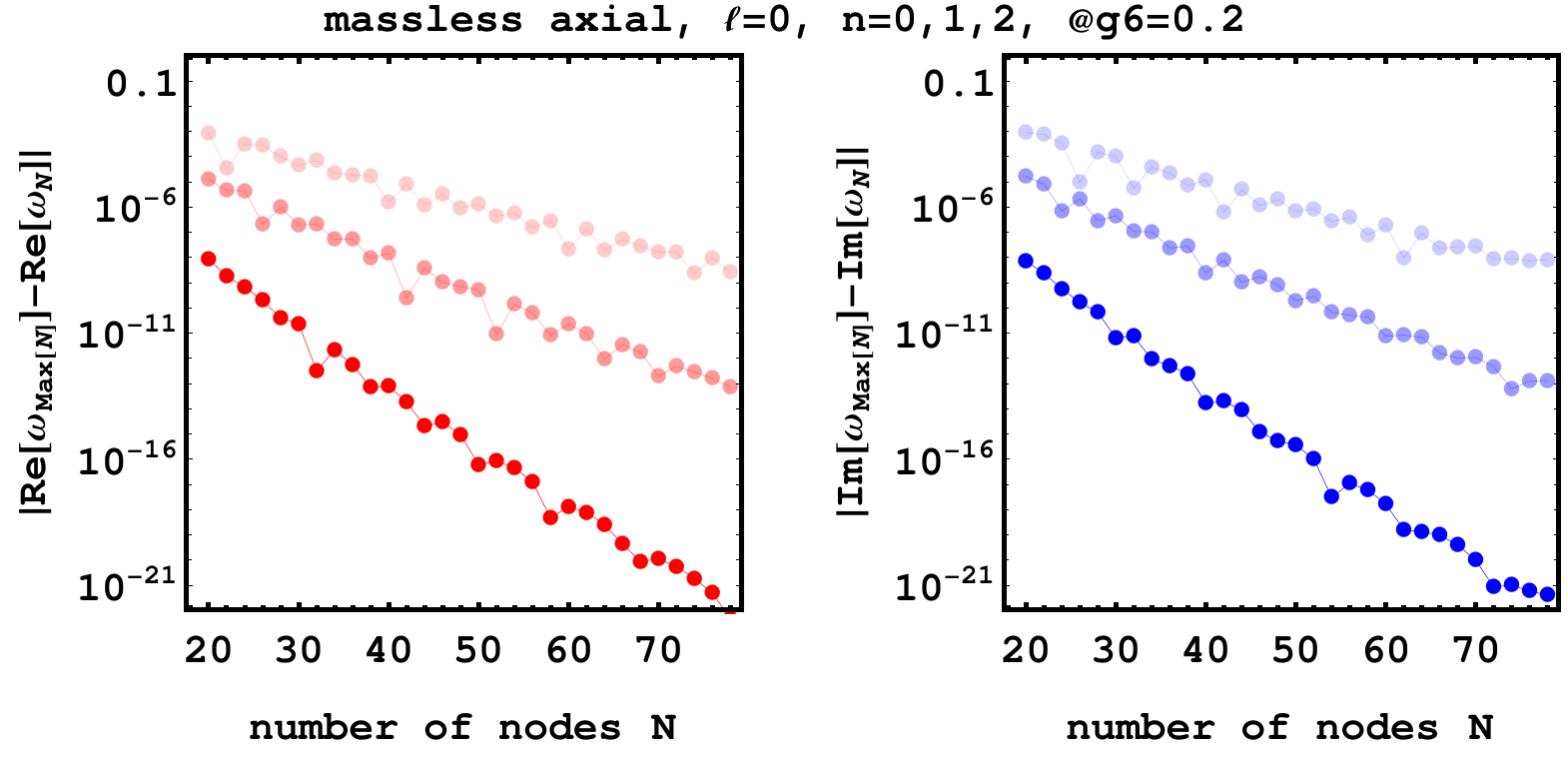}
	\hfill
	\includegraphics[width=0.485\textwidth]{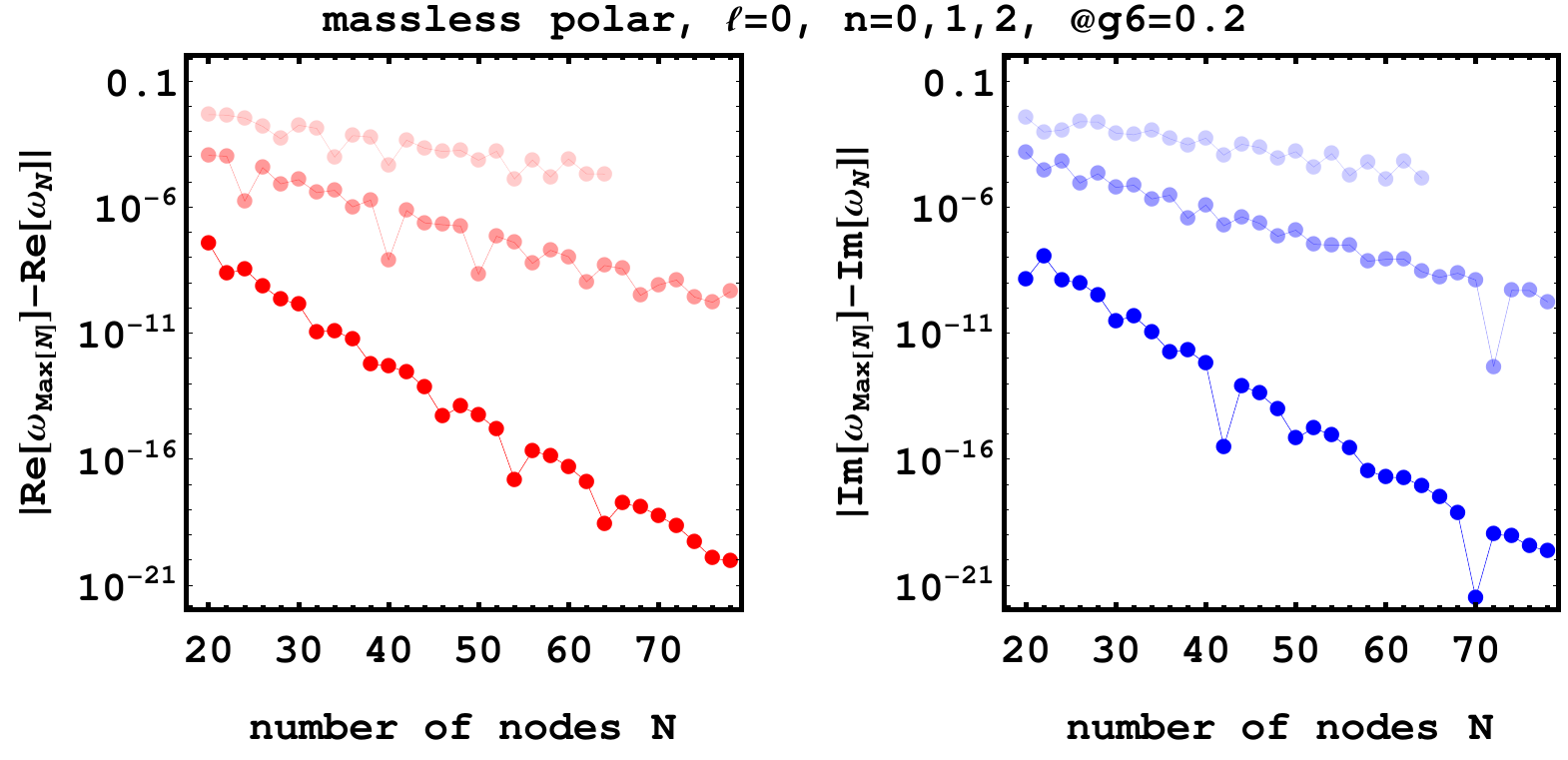}
	\\
	\includegraphics[width=0.485\textwidth]{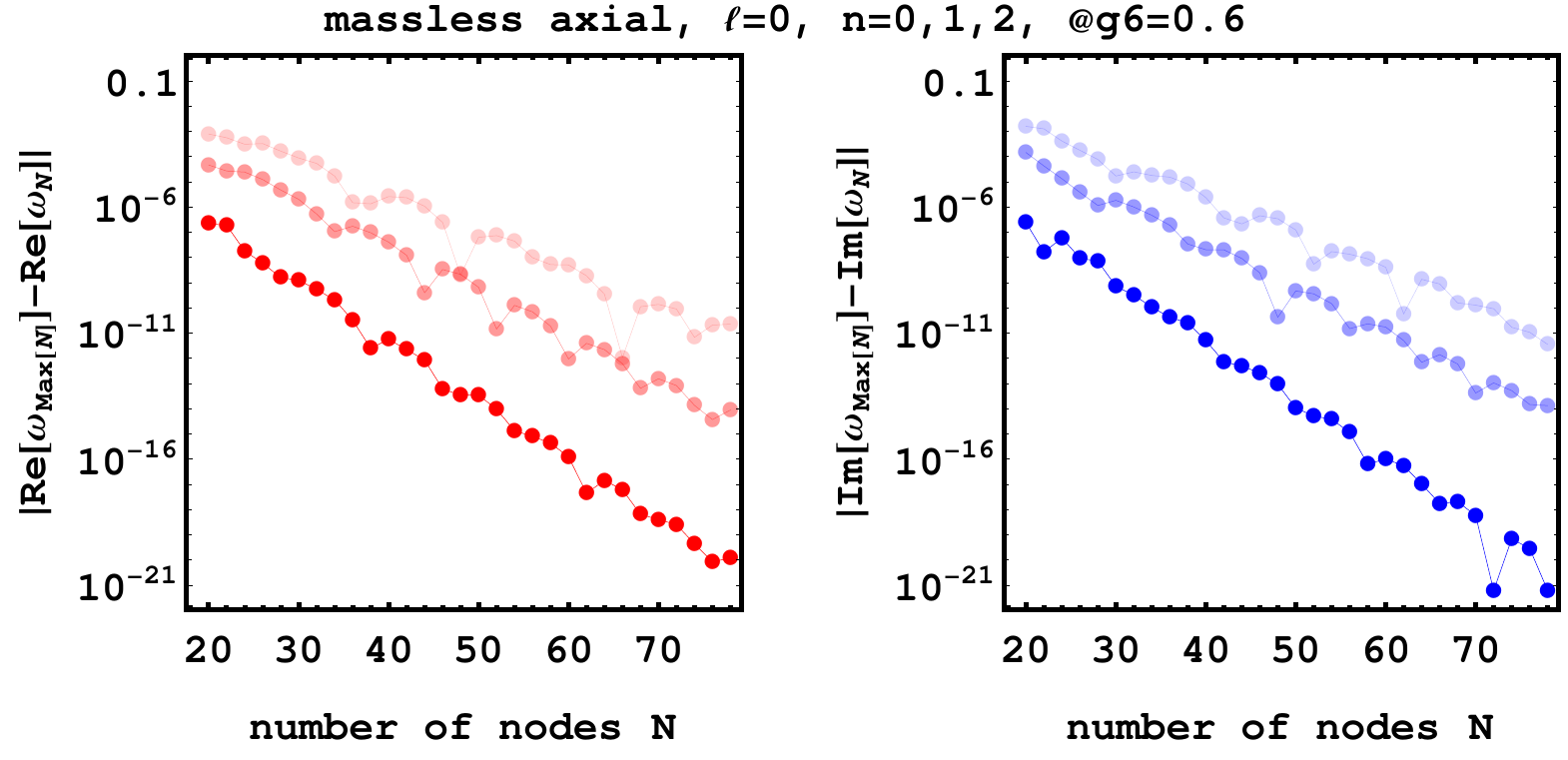}
	\hfill
	\includegraphics[width=0.485\textwidth]{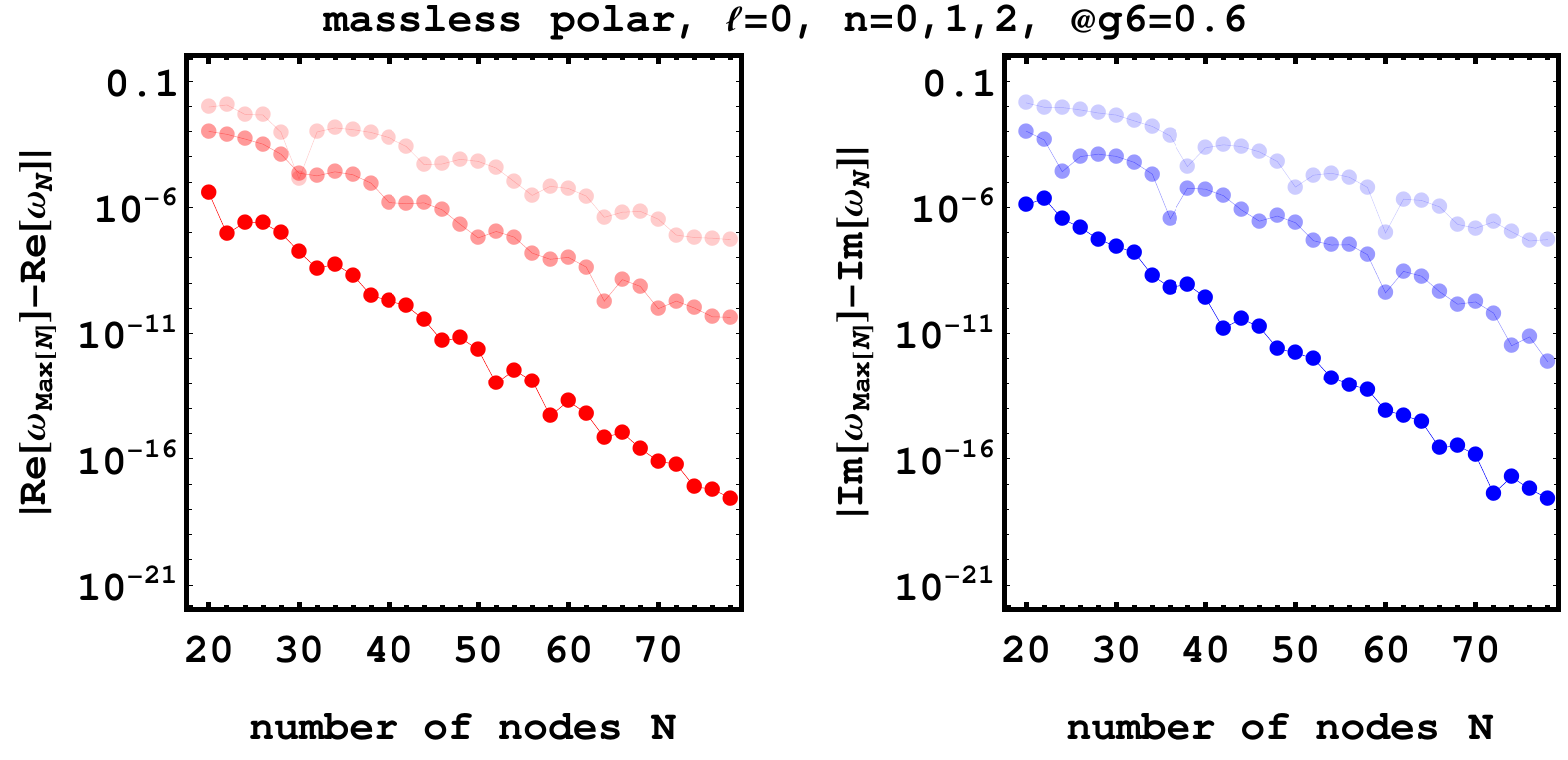}
	\caption{Exponential convergence with growing number $N$ of Chebyshev nodes for the massless axial and polar modes. The left two columns show real and imaginary parts of the massless axial mode. The right two columns show real and imaginary parts of the massless polar mode. From top to bottom, we show convergence plots for exemplary points at $g_6=-0.2,\,0.2,\,0.6$. As in Fig.~\ref{fig:broken-isospectrality-massless}, increasingly light shading denotes increasing $n$. To aid direct visual comparison, we choose the same plot range on the y-axis.
}
	\label{fig:convergence}
\end{figure}

For all the presented figures we have performed accuracy checks (i) at the points closest to the respective poles at $g_6=-1/2$ and $g_6=1$ as well as (ii) at other random points. For the fundamental modes, these converence tests agree very well with expectations from an analytical error estimate discussed, for instance, in \cite[App.~C.4]{Baumann:2019eav}. In particular, we find exponential convergence with growing number $N$ of Chebyshev nodes. Moreover, we can also see that the convergence properties worsen with closeness to singularities in the complex plane.

We also find that higher modes ($n>1$) beyond the fundamental ($n=1$) are increasingly difficult to obtain because convergence worsens significantly. While we do not provide an analytical argument, we expect that the underlying reason is an increase in the number of oscillations (in the radial coordinate $r$) with growing $n$. The more oscillatory the behaviour, the harder it becomes to resolve these oscillations with fixed number of nodes $N$.

The $n=1$ and $n=2$ results for the massless modes, cf.~Fig.~\ref{fig:broken-isospectrality-massless} in the main text, are thus numerically most challenging to obtain. In Fig.~\ref{fig:convergence}, we present the respective convergence plots for exemplary values of $g_6$.

\subsection{Values of QNM frequencies}

We provide the numerical values of the frequencies in Tables.~\ref{tab:mono_qnm_n0}-\ref{tab:polar-vector_qnm_n0} for readers to make comparison.

\begin{table}[tbp]
\centering
\footnotesize
\begin{tabular}{|l|c|c|c|c|c|}
\hline
\diagbox{$\mu r_g$}{$g_6$} & $-0.25$ & $0$ & $0.25$ & $0.5$ & $0.75$\\
\hline 
$0.1$ & $0.2246 -0.1971 i$ & $0.2247 -0.1972 i$ & $0.2247 -0.1972 i$ & $0.2245 -0.1972 i$ & $0.2244 -0.1971 i$\\
\hline
$0.2$ & $0.2430 -0.1575 i$ & $0.2432 -0.1582 i$ & $0.2427 -0.1582 i$ & $0.2421-0.1580 i$ & $0.2414 -0.1576 i$\\
\hline
$0.3$ & $0.2941 -0.1057 i$ & $0.2929 -0.1061 i$ & $0.2916 -0.1054 i$ & $0.2905 -0.1043 i$ & $0.2895 -0.1031 i$\\
\hline
$0.4$ & $0.3659 -0.0640 i$ & $0.3643 -0.0628 i$ & $0.3632 -0.0607 i$ & $0.3623-0.0585 i$ & $0.3617 -0.0563 i$\\
\hline
\end{tabular}
\caption{\label{tab:mono_qnm_n0} Frequencies $\omega r_g$ of the fundamental $(n=0)$ monopole QNMs.}
\end{table}
\begin{table}[tbp]
\centering
\footnotesize
\begin{tabular}{|l|c|c|c|c|c|}
\hline
\diagbox{$\mu r_g$}{$g_6$} & $-0.25$ & $0$ & $0.25$ & $0.5$ & $0.75$\\
\hline 
$0.1$ & $0.4451 -0.1926 i$ & $0.4994 -0.1831 i$ & $0.5718 -0.1861 i$ & $0.6748 -0.2017 i$ & $0.8525 -0.2249 i$\\
\hline
$0.2$ & $0.4563 -0.1847 i$ & $0.5079 -0.1774 i$ & $0.5780 -0.1818 i$ & $0.6792-0.1987 i$ & $0.8559 -0.2236 i$\\
\hline
$0.3$ & $0.4747 -0.1717 i$ & $0.5222 -0.1679 i$ & $0.5884 -0.1744 i$ & $0.6865 -0.1937 i$ & $0.8617 -0.2214 i$\\
\hline
$0.4$ & $0.5005 -0.1537 i$ & $0.5424 -0.1542 i$ & $0.6032 -0.1637 i$ & $0.6966 -0.1863 i$ & $0.8696 -0.2182 i$\\
\hline
\end{tabular}
\caption{\label{tab:axial_qnm_n0} Frequencies $\omega r_g$ of the fundamental $(n=0)$, first multipole $(\ell=1)$ axial QNMs.}
\end{table}
\begin{table}[tbp]
\centering
\footnotesize
\begin{tabular}{|l|c|c|c|c|c|}
\hline
\diagbox{$\mu r_g$}{$g_6$} & $-0.25$ & $0$ & $0.25$ & $0.5$ & $0.75$\\
\hline 
$0.1$ & $0.6082 -0.1854 i$ & $0.5925 -0.1916 i$ & $0.591 -0.1924 i$ & $0.5904 \
-0.1925 i$ & $0.59 -0.1925 i$\\
\hline
$0.2$ & $0.6347 -0.1775 i$ & $0.6107 -0.1829 i$ & $0.6059 -0.1844 i$ & \
$0.6037 -0.1847 i$ & $0.6023 -0.1847 i$\\
\hline
$0.3$ & $0.6644 -0.1682 i$ & $0.6374 -0.172 i$ & $0.6295 -0.1732 i$ & $0.6254 \
-0.1731 i$ & $0.6227 -0.1728 i$\\
\hline
$0.4$ & $0.6977 -0.1577 i$ & $0.6703 -0.1599 i$ & $0.6603 -0.16 i$ & $0.6547 \
-0.1592 i$ & $0.651 -0.1582 i$\\
\hline
\end{tabular}
\caption{\label{tab:polar-scalar_qnm_n0} Frequencies $\omega r_g$ of the fundamental $(n=0)$, first multipole $(\ell=1)$ polar (scalar) QNMs.}
\end{table}
\begin{table}[tbp]
\centering
\footnotesize
\begin{tabular}{|l|c|c|c|c|c|}
\hline
\diagbox{$\mu r_g$}{$g_6$} & $-0.25$ & $0$ & $0.25$ & $0.5$ & $0.75$\\
\hline 
$0.1$ & $0.5661 -0.1966 i$ & $0.4938 -0.1865 i$ & $0.4485 -0.196 i$ & $0.4374 -0.2133 i$ & $0.4596 -0.2112 i$\\
\hline
$0.2$ & $0.5492 -0.1994 i$ & $0.487 -0.1887 i$ & $0.4462 -0.1969 i$ & $0.4375 -0.2136 i$ & $0.4617 -0.2115 i$\\
\hline
$0.3$ & $0.5349 -0.1999 i$ & $0.4792 -0.1885 i$ & $0.4431 -0.1960 i$ & $0.4374 -0.2128 i$ & $0.4647 -0.2114 i$\\
\hline
$0.4$ & $0.5229 -0.1979 i$ & $0.4722 -0.1851 i$ & $0.44 -0.1923 i$ & $0.4366 -0.2101 i$ & $0.4679 -0.211 i$\\
\hline
\end{tabular}
\caption{\label{tab:polar-vector_qnm_n0} Frequencies $\omega r_g$ of the fundamental $(n=0)$, first multipole $(\ell=1)$ polar (vector) QNMs.}
\end{table}

\bigskip


\bibliographystyle{JHEP}
\bibliography{refs}

\providecommand{\href}[2]{#2}\begingroup\raggedright\begin{thebibliography}{10}

\bibitem{LIGOScientific:2016aoc}
{\scshape LIGO Scientific, Virgo} collaboration, \emph{{Observation of
  Gravitational Waves from a Binary Black Hole Merger}},
  \href{https://doi.org/10.1103/PhysRevLett.116.061102}{\emph{Phys. Rev. Lett.}
  {\bfseries 116} (2016) 061102}
  [\href{https://arxiv.org/abs/1602.03837}{{\ttfamily 1602.03837}}].

\bibitem{EventHorizonTelescope:2019dse}
{\scshape Event Horizon Telescope} collaboration, \emph{{First M87 Event
  Horizon Telescope Results. I. The Shadow of the Supermassive Black Hole}},
  \href{https://doi.org/10.3847/2041-8213/ab0ec7}{\emph{Astrophys. J. Lett.}
  {\bfseries 875} (2019) L1}
  [\href{https://arxiv.org/abs/1906.11238}{{\ttfamily 1906.11238}}].

\bibitem{Regge:1957td}
T.~Regge and J.~A. Wheeler, \emph{{Stability of a Schwarzschild singularity}},
  \href{https://doi.org/10.1103/PhysRev.108.1063}{\emph{Phys. Rev.} {\bfseries
  108} (1957) 1063}.

\bibitem{Zerilli:1970se}
F.~J. Zerilli, \emph{{Effective potential for even parity Regge-Wheeler
  gravitational perturbation equations}},
  \href{https://doi.org/10.1103/PhysRevLett.24.737}{\emph{Phys. Rev. Lett.}
  {\bfseries 24} (1970) 737}.

\bibitem{Zerilli:1970wzz}
F.~J. Zerilli, \emph{{Gravitational field of a particle falling in a
  schwarzschild geometry analyzed in tensor harmonics}},
  \href{https://doi.org/10.1103/PhysRevD.2.2141}{\emph{Phys. Rev. D} {\bfseries
  2} (1970) 2141}.

\bibitem{Moncrief:1974am}
V.~Moncrief, \emph{{Gravitational perturbations of spherically symmetric
  systems. I. The exterior problem.}},
  \href{https://doi.org/10.1016/0003-4916(74)90173-0}{\emph{Annals Phys.}
  {\bfseries 88} (1974) 323}.

\bibitem{Chandrasekhar:1975zza}
S.~Chandrasekhar and S.~L. Detweiler, \emph{{The quasi-normal modes of the
  Schwarzschild black hole}},
  \href{https://doi.org/10.1098/rspa.1975.0112}{\emph{Proc. Roy. Soc. Lond. A}
  {\bfseries 344} (1975) 441}.

\bibitem{Teukolsky:1972my}
S.~A. Teukolsky, \emph{{Rotating black holes - separable wave equations for
  gravitational and electromagnetic perturbations}},
  \href{https://doi.org/10.1103/PhysRevLett.29.1114}{\emph{Phys. Rev. Lett.}
  {\bfseries 29} (1972) 1114}.

\bibitem{Moncrief:1974gw}
V.~Moncrief, \emph{{Odd-parity stability of a Reissner-Nordstrom black hole}},
  \href{https://doi.org/10.1103/PhysRevD.9.2707}{\emph{Phys. Rev. D} {\bfseries
  9} (1974) 2707}.

\bibitem{Moncrief:1974ng}
V.~Moncrief, \emph{{Stability of Reissner-Nordstrom black holes}},
  \href{https://doi.org/10.1103/PhysRevD.10.1057}{\emph{Phys. Rev. D}
  {\bfseries 10} (1974) 1057}.

\bibitem{Nollert:1999ji}
H.-P. Nollert, \emph{{TOPICAL REVIEW: Quasinormal modes: the characteristic
  `sound' of black holes and neutron stars}},
  \href{https://doi.org/10.1088/0264-9381/16/12/201}{\emph{Class. Quant. Grav.}
  {\bfseries 16} (1999) R159}.

\bibitem{Kokkotas:1999bd}
K.~D. Kokkotas and B.~G. Schmidt, \emph{{Quasinormal modes of stars and black
  holes}}, \href{https://doi.org/10.12942/lrr-1999-2}{\emph{Living Rev. Rel.}
  {\bfseries 2} (1999) 2}
  [\href{https://arxiv.org/abs/gr-qc/9909058}{{\ttfamily gr-qc/9909058}}].

\bibitem{Berti:2009kk}
E.~Berti, V.~Cardoso and A.~O. Starinets, \emph{{Quasinormal modes of black
  holes and black branes}},
  \href{https://doi.org/10.1088/0264-9381/26/16/163001}{\emph{Class. Quant.
  Grav.} {\bfseries 26} (2009) 163001}
  [\href{https://arxiv.org/abs/0905.2975}{{\ttfamily 0905.2975}}].

\bibitem{Konoplya:2011qq}
R.~A. Konoplya and A.~Zhidenko, \emph{{Quasinormal modes of black holes: From
  astrophysics to string theory}},
  \href{https://doi.org/10.1103/RevModPhys.83.793}{\emph{Rev. Mod. Phys.}
  {\bfseries 83} (2011) 793} [\href{https://arxiv.org/abs/1102.4014}{{\ttfamily
  1102.4014}}].

\bibitem{Berti:2018vdi}
E.~Berti, K.~Yagi, H.~Yang and N.~Yunes, \emph{{Extreme Gravity Tests with
  Gravitational Waves from Compact Binary Coalescences: (II) Ringdown}},
  \href{https://doi.org/10.1007/s10714-018-2372-6}{\emph{Gen. Rel. Grav.}
  {\bfseries 50} (2018) 49} [\href{https://arxiv.org/abs/1801.03587}{{\ttfamily
  1801.03587}}].

\bibitem{Cardoso:2018ptl}
V.~Cardoso, M.~Kimura, A.~Maselli and L.~Senatore, \emph{{Black Holes in an
  Effective Field Theory Extension of General Relativity}},
  \href{https://doi.org/10.1103/PhysRevLett.121.251105}{\emph{Phys. Rev. Lett.}
  {\bfseries 121} (2018) 251105}
  [\href{https://arxiv.org/abs/1808.08962}{{\ttfamily 1808.08962}}].

\bibitem{Franciolini:2018uyq}
G.~Franciolini, L.~Hui, R.~Penco, L.~Santoni and E.~Trincherini,
  \emph{{Effective Field Theory of Black Hole Quasinormal Modes in
  Scalar-Tensor Theories}},
  \href{https://doi.org/10.1007/JHEP02(2019)127}{\emph{JHEP} {\bfseries 02}
  (2019) 127} [\href{https://arxiv.org/abs/1810.07706}{{\ttfamily
  1810.07706}}].

\bibitem{deRham:2020ejn}
C.~de~Rham, J.~Francfort and J.~Zhang, \emph{{Black Hole Gravitational Waves in
  the Effective Field Theory of Gravity}},
  \href{https://doi.org/10.1103/PhysRevD.102.024079}{\emph{Phys. Rev. D}
  {\bfseries 102} (2020) 024079}
  [\href{https://arxiv.org/abs/2005.13923}{{\ttfamily 2005.13923}}].

\bibitem{Cano:2021myl}
P.~A. Cano, K.~Fransen, T.~Hertog and S.~Maenaut, \emph{{Gravitational ringing
  of rotating black holes in higher-derivative gravity}},
  \href{https://arxiv.org/abs/2110.11378}{{\ttfamily 2110.11378}}.

\bibitem{Cardoso:2009pk}
V.~Cardoso and L.~Gualtieri, \emph{{Perturbations of Schwarzschild black holes
  in Dynamical Chern-Simons modified gravity}},
  \href{https://doi.org/10.1103/PhysRevD.81.089903}{\emph{Phys. Rev. D}
  {\bfseries 80} (2009) 064008}
  [\href{https://arxiv.org/abs/0907.5008}{{\ttfamily 0907.5008}}].

\bibitem{Molina:2010fb}
C.~Molina, P.~Pani, V.~Cardoso and L.~Gualtieri, \emph{{Gravitational signature
  of Schwarzschild black holes in dynamical Chern-Simons gravity}},
  \href{https://doi.org/10.1103/PhysRevD.81.124021}{\emph{Phys. Rev. D}
  {\bfseries 81} (2010) 124021}
  [\href{https://arxiv.org/abs/1004.4007}{{\ttfamily 1004.4007}}].

\bibitem{Tattersall:2018nve}
O.~J. Tattersall and P.~G. Ferreira, \emph{{Quasinormal modes of black holes in
  Horndeski gravity}},
  \href{https://doi.org/10.1103/PhysRevD.97.104047}{\emph{Phys. Rev. D}
  {\bfseries 97} (2018) 104047}
  [\href{https://arxiv.org/abs/1804.08950}{{\ttfamily 1804.08950}}].

\bibitem{Wagle:2021tam}
P.~Wagle, N.~Yunes and H.~O. Silva, \emph{{Quasinormal modes of slowly-rotating
  black holes in dynamical Chern-Simons gravity}},
  \href{https://arxiv.org/abs/2103.09913}{{\ttfamily 2103.09913}}.

\bibitem{Bryant:2021xdh}
A.~Bryant, H.~O. Silva, K.~Yagi and K.~Glampedakis, \emph{{Eikonal quasinormal
  modes of black holes beyond general relativity. III. Scalar Gauss-Bonnet
  gravity}}, \href{https://doi.org/10.1103/PhysRevD.104.044051}{\emph{Phys.
  Rev. D} {\bfseries 104} (2021) 044051}
  [\href{https://arxiv.org/abs/2106.09657}{{\ttfamily 2106.09657}}].

\bibitem{Pierini:2021jxd}
L.~Pierini and L.~Gualtieri, \emph{{Quasi-normal modes of rotating black holes
  in Einstein-dilaton Gauss-Bonnet gravity: the first order in rotation}},
  \href{https://doi.org/10.1103/PhysRevD.103.124017}{\emph{Phys. Rev. D}
  {\bfseries 103} (2021) 124017}
  [\href{https://arxiv.org/abs/2103.09870}{{\ttfamily 2103.09870}}].

\bibitem{Blazquez-Salcedo:2016enn}
J.~L. Bl\'azquez-Salcedo, C.~F.~B. Macedo, V.~Cardoso, V.~Ferrari,
  L.~Gualtieri, F.~S. Khoo et~al., \emph{{Perturbed black holes in
  Einstein-dilaton-Gauss-Bonnet gravity: Stability, ringdown, and
  gravitational-wave emission}},
  \href{https://doi.org/10.1103/PhysRevD.94.104024}{\emph{Phys. Rev. D}
  {\bfseries 94} (2016) 104024}
  [\href{https://arxiv.org/abs/1609.01286}{{\ttfamily 1609.01286}}].

\bibitem{Chandrasekhar:1985kt}
S.~Chandrasekhar, \emph{The mathematical theory of black holes}. Oxford
  University Press, New York, 1983.

\bibitem{Galtsov:1984ixy}
D.~V. Gal'tsov, G.~V. Pomerantseva and G.~A. Chizhov, \emph{{Behavior of
  massive vector particles in a Schwarzschild field}},
  \href{https://doi.org/10.1007/BF00893117}{\emph{Sov. Phys. J.} {\bfseries 27}
  (1984) 697}.

\bibitem{Konoplya:2005hr}
R.~A. Konoplya, \emph{{Massive vector field perturbations in the Schwarzschild
  background: Stability and unusual quasinormal spectrum}},
  \href{https://doi.org/10.1103/PhysRevD.73.024009}{\emph{Phys. Rev. D}
  {\bfseries 73} (2006) 024009}
  [\href{https://arxiv.org/abs/gr-qc/0509026}{{\ttfamily gr-qc/0509026}}].

\bibitem{Rosa:2011my}
J.~G. Rosa and S.~R. Dolan, \emph{{Massive vector fields on the Schwarzschild
  spacetime: quasi-normal modes and bound states}},
  \href{https://doi.org/10.1103/PhysRevD.85.044043}{\emph{Phys. Rev.}
  {\bfseries D85} (2012) 044043}
  [\href{https://arxiv.org/abs/1110.4494}{{\ttfamily 1110.4494}}].

\bibitem{Fernandes:2021qvr}
T.~V. Fernandes, D.~Hilditch, J.~P.~S. Lemos and V.~Cardoso, \emph{{Quasinormal
  modes of Proca fields in Schwarzschild-AdS spacetime}},
  \href{https://arxiv.org/abs/2112.03282}{{\ttfamily 2112.03282}}.

\bibitem{Frolov:2018ezx}
V.~P. Frolov, P.~Krtou\v{s}, D.~Kubiz\v{n}\'ak and J.~E. Santos, \emph{{Massive
  Vector Fields in Rotating Black-Hole Spacetimes: Separability and Quasinormal
  Modes}}, \href{https://doi.org/10.1103/PhysRevLett.120.231103}{\emph{Phys.
  Rev. Lett.} {\bfseries 120} (2018) 231103}
  [\href{https://arxiv.org/abs/1804.00030}{{\ttfamily 1804.00030}}].

\bibitem{Baumann:2019eav}
D.~Baumann, H.~S. Chia, J.~Stout and L.~ter Haar, \emph{{The Spectra of
  Gravitational Atoms}},
  \href{https://doi.org/10.1088/1475-7516/2019/12/006}{\emph{JCAP} {\bfseries
  12} (2019) 006} [\href{https://arxiv.org/abs/1908.10370}{{\ttfamily
  1908.10370}}].

\bibitem{Percival:2020skc}
J.~Percival and S.~R. Dolan, \emph{{Quasinormal modes of massive vector fields
  on the Kerr spacetime}},
  \href{https://doi.org/10.1103/PhysRevD.102.104055}{\emph{Phys. Rev. D}
  {\bfseries 102} (2020) 104055}
  [\href{https://arxiv.org/abs/2008.10621}{{\ttfamily 2008.10621}}].

\bibitem{Chagoya:2017ojn}
J.~Chagoya and G.~Tasinato, \emph{{Stealth configurations in vector-tensor
  theories of gravity}},
  \href{https://doi.org/10.1088/1475-7516/2018/01/046}{\emph{JCAP} {\bfseries
  01} (2018) 046} [\href{https://arxiv.org/abs/1707.07951}{{\ttfamily
  1707.07951}}].

\bibitem{Tasinato:2014eka}
G.~Tasinato, \emph{{Cosmic Acceleration from Abelian Symmetry Breaking}},
  \href{https://doi.org/10.1007/JHEP04(2014)067}{\emph{JHEP} {\bfseries 04}
  (2014) 067} [\href{https://arxiv.org/abs/1402.6450}{{\ttfamily 1402.6450}}].

\bibitem{Heisenberg:2014rta}
L.~Heisenberg, \emph{{Generalization of the Proca Action}},
  \href{https://doi.org/10.1088/1475-7516/2014/05/015}{\emph{JCAP} {\bfseries
  05} (2014) 015} [\href{https://arxiv.org/abs/1402.7026}{{\ttfamily
  1402.7026}}].

\bibitem{Jimenez:2013qsa}
J.~Beltran~Jimenez, R.~Durrer, L.~Heisenberg and M.~Thorsrud, \emph{{Stability
  of Horndeski vector-tensor interactions}},
  \href{https://doi.org/10.1088/1475-7516/2013/10/064}{\emph{JCAP} {\bfseries
  1310} (2013) 064} [\href{https://arxiv.org/abs/1308.1867}{{\ttfamily
  1308.1867}}].

\bibitem{Garcia-Saenz:2021uyv}
S.~Garcia-Saenz, A.~Held and J.~Zhang, \emph{{Destabilization of Black Holes
  and Stars by Generalized Proca Fields}},
  \href{https://doi.org/10.1103/PhysRevLett.127.131104}{\emph{Phys. Rev. Lett.}
  {\bfseries 127} (2021) 131104}
  [\href{https://arxiv.org/abs/2104.08049}{{\ttfamily 2104.08049}}].

\bibitem{Chandrasekhar:1975nkd}
S.~Chandrasekhar, \emph{{On the equations governing the perturbations of the
  Schwarzschild black hole}},
  \href{https://doi.org/10.1098/rspa.1975.0066}{\emph{Proc. Roy. Soc. Lond. A}
  {\bfseries 343} (1975) 289}.

\bibitem{Brito:2013yxa}
R.~Brito, V.~Cardoso and P.~Pani, \emph{{Partially massless gravitons do not
  destroy general relativity black holes}},
  \href{https://doi.org/10.1103/PhysRevD.87.124024}{\emph{Phys. Rev. D}
  {\bfseries 87} (2013) 124024}
  [\href{https://arxiv.org/abs/1306.0908}{{\ttfamily 1306.0908}}].

\bibitem{Rosen:2020crj}
R.~A. Rosen and L.~Santoni, \emph{{Black hole perturbations of massive and
  partially massless spin-2 fields in (anti) de Sitter spacetime}},
  \href{https://doi.org/10.1007/JHEP03(2021)139}{\emph{JHEP} {\bfseries 03}
  (2021) 139} [\href{https://arxiv.org/abs/2010.00595}{{\ttfamily
  2010.00595}}].

\bibitem{Konoplya:2003dd}
R.~A. Konoplya, \emph{{Gravitational quasinormal radiation of higher
  dimensional black holes}},
  \href{https://doi.org/10.1103/PhysRevD.68.124017}{\emph{Phys. Rev. D}
  {\bfseries 68} (2003) 124017}
  [\href{https://arxiv.org/abs/hep-th/0309030}{{\ttfamily hep-th/0309030}}].

\bibitem{Cardoso:2001bb}
V.~Cardoso and J.~P.~S. Lemos, \emph{{Quasinormal modes of Schwarzschild
  anti-de Sitter black holes: Electromagnetic and gravitational
  perturbations}},
  \href{https://doi.org/10.1103/PhysRevD.64.084017}{\emph{Phys. Rev. D}
  {\bfseries 64} (2001) 084017}
  [\href{https://arxiv.org/abs/gr-qc/0105103}{{\ttfamily gr-qc/0105103}}].

\bibitem{Brito:2013wya}
R.~Brito, V.~Cardoso and P.~Pani, \emph{{Massive spin-2 fields on black hole
  spacetimes: Instability of the Schwarzschild and Kerr solutions and bounds on
  the graviton mass}},
  \href{https://doi.org/10.1103/PhysRevD.88.023514}{\emph{Phys. Rev. D}
  {\bfseries 88} (2013) 023514}
  [\href{https://arxiv.org/abs/1304.6725}{{\ttfamily 1304.6725}}].

\bibitem{Chaverra:2016ttw}
E.~Chaverra, J.~C. Degollado, C.~Moreno and O.~Sarbach, \emph{{Black holes in
  nonlinear electrodynamics: Quasinormal spectra and parity splitting}},
  \href{https://doi.org/10.1103/PhysRevD.93.123013}{\emph{Phys. Rev. D}
  {\bfseries 93} (2016) 123013}
  [\href{https://arxiv.org/abs/1605.04003}{{\ttfamily 1605.04003}}].

\bibitem{Nomura:2021efi}
K.~Nomura and D.~Yoshida, \emph{{Quasinormal modes of charged black holes with
  corrections from nonlinear electrodynamics}},
  \href{https://arxiv.org/abs/2111.06273}{{\ttfamily 2111.06273}}.

\bibitem{Love1912}
A.~E.~H. Love, \emph{{Some Problems of Geodynamics}},
  \href{https://doi.org/10.1038/089471a0}{\emph{Nature} {\bfseries 89} (1912)
  471–472}.

\bibitem{Flanagan:2007ix}
E.~E. Flanagan and T.~Hinderer, \emph{{Constraining neutron star tidal Love
  numbers with gravitational wave detectors}},
  \href{https://doi.org/10.1103/PhysRevD.77.021502}{\emph{Phys. Rev. D}
  {\bfseries 77} (2008) 021502}
  [\href{https://arxiv.org/abs/0709.1915}{{\ttfamily 0709.1915}}].

\bibitem{Damour:2009vw}
T.~Damour and A.~Nagar, \emph{{Relativistic tidal properties of neutron
  stars}}, \href{https://doi.org/10.1103/PhysRevD.80.084035}{\emph{Phys. Rev.
  D} {\bfseries 80} (2009) 084035}
  [\href{https://arxiv.org/abs/0906.0096}{{\ttfamily 0906.0096}}].

\bibitem{Binnington:2009bb}
T.~Binnington and E.~Poisson, \emph{{Relativistic theory of tidal Love
  numbers}}, \href{https://doi.org/10.1103/PhysRevD.80.084018}{\emph{Phys. Rev.
  D} {\bfseries 80} (2009) 084018}
  [\href{https://arxiv.org/abs/0906.1366}{{\ttfamily 0906.1366}}].

\bibitem{LeTiec:2020spy}
A.~Le~Tiec and M.~Casals, \emph{{Spinning Black Holes Fall in Love}},
  \href{https://doi.org/10.1103/PhysRevLett.126.131102}{\emph{Phys. Rev. Lett.}
  {\bfseries 126} (2021) 131102}
  [\href{https://arxiv.org/abs/2007.00214}{{\ttfamily 2007.00214}}].

\bibitem{Chia:2020yla}
H.~S. Chia, \emph{{Tidal deformation and dissipation of rotating black holes}},
  \href{https://doi.org/10.1103/PhysRevD.104.024013}{\emph{Phys. Rev. D}
  {\bfseries 104} (2021) 024013}
  [\href{https://arxiv.org/abs/2010.07300}{{\ttfamily 2010.07300}}].

\bibitem{Goldberger:2020fot}
W.~D. Goldberger, J.~Li and I.~Z. Rothstein, \emph{{Non-conservative effects on
  spinning black holes from world-line effective field theory}},
  \href{https://doi.org/10.1007/JHEP06(2021)053}{\emph{JHEP} {\bfseries 06}
  (2021) 053} [\href{https://arxiv.org/abs/2012.14869}{{\ttfamily
  2012.14869}}].

\bibitem{Hui:2020xxx}
L.~Hui, A.~Joyce, R.~Penco, L.~Santoni and A.~R. Solomon, \emph{{Static
  response and Love numbers of Schwarzschild black holes}},
  \href{https://doi.org/10.1088/1475-7516/2021/04/052}{\emph{JCAP} {\bfseries
  04} (2021) 052} [\href{https://arxiv.org/abs/2010.00593}{{\ttfamily
  2010.00593}}].

\bibitem{LeTiec:2020bos}
A.~Le~Tiec, M.~Casals and E.~Franzin, \emph{{Tidal Love Numbers of Kerr Black
  Holes}}, \href{https://doi.org/10.1103/PhysRevD.103.084021}{\emph{Phys. Rev.
  D} {\bfseries 103} (2021) 084021}
  [\href{https://arxiv.org/abs/2010.15795}{{\ttfamily 2010.15795}}].

\bibitem{Charalambous:2021mea}
P.~Charalambous, S.~Dubovsky and M.~M. Ivanov, \emph{{On the Vanishing of Love
  Numbers for Kerr Black Holes}},
  \href{https://doi.org/10.1007/JHEP05(2021)038}{\emph{JHEP} {\bfseries 05}
  (2021) 038} [\href{https://arxiv.org/abs/2102.08917}{{\ttfamily
  2102.08917}}].

\bibitem{Pereniguez:2021xcj}
D.~Pere\~niguez and V.~Cardoso, \emph{{Love numbers and magnetic susceptibility
  of charged black holes}},  \href{https://arxiv.org/abs/2112.08400}{{\ttfamily
  2112.08400}}.

\bibitem{Kol:2011vg}
B.~Kol and M.~Smolkin, \emph{{Black hole stereotyping: Induced gravito-static
  polarization}}, \href{https://doi.org/10.1007/JHEP02(2012)010}{\emph{JHEP}
  {\bfseries 02} (2012) 010} [\href{https://arxiv.org/abs/1110.3764}{{\ttfamily
  1110.3764}}].

\bibitem{Cardoso:2019vof}
V.~Cardoso, L.~Gualtieri and C.~J. Moore, \emph{{Gravitational waves and higher
  dimensions: Love numbers and Kaluza-Klein excitations}},
  \href{https://doi.org/10.1103/PhysRevD.100.124037}{\emph{Phys. Rev. D}
  {\bfseries 100} (2019) 124037}
  [\href{https://arxiv.org/abs/1910.09557}{{\ttfamily 1910.09557}}].

\bibitem{Cardoso:2017cfl}
V.~Cardoso, E.~Franzin, A.~Maselli, P.~Pani and G.~Raposo, \emph{{Testing
  strong-field gravity with tidal Love numbers}},
  \href{https://doi.org/10.1103/PhysRevD.95.084014}{\emph{Phys. Rev. D}
  {\bfseries 95} (2017) 084014}
  [\href{https://arxiv.org/abs/1701.01116}{{\ttfamily 1701.01116}}].

\bibitem{Cai:2019npx}
S.~Cai and K.-D. Wang, \emph{{Non-vanishing of tidal Love numbers}},
  \href{https://arxiv.org/abs/1906.06850}{{\ttfamily 1906.06850}}.

\bibitem{deRham:2016wji}
C.~de~Rham and A.~Matas, \emph{{Ostrogradsky in Theories with Multiple
  Fields}}, \href{https://doi.org/10.1088/1475-7516/2016/06/041}{\emph{JCAP}
  {\bfseries 06} (2016) 041}
  [\href{https://arxiv.org/abs/1604.08638}{{\ttfamily 1604.08638}}].

\bibitem{Horndeski:1976gi}
G.~W. Horndeski, \emph{{Conservation of Charge and the Einstein-Maxwell Field
  Equations}}, \href{https://doi.org/10.1063/1.522837}{\emph{J. Math. Phys.}
  {\bfseries 17} (1976) 1980}.

\bibitem{Allys:2015sht}
E.~Allys, P.~Peter and Y.~Rodriguez, \emph{{Generalized Proca action for an
  Abelian vector field}},
  \href{https://doi.org/10.1088/1475-7516/2016/02/004}{\emph{JCAP} {\bfseries
  02} (2016) 004} [\href{https://arxiv.org/abs/1511.03101}{{\ttfamily
  1511.03101}}].

\bibitem{BeltranJimenez:2016rff}
J.~Beltran~Jimenez and L.~Heisenberg, \emph{{Derivative self-interactions for a
  massive vector field}},
  \href{https://doi.org/10.1016/j.physletb.2016.04.017}{\emph{Phys. Lett. B}
  {\bfseries 757} (2016) 405}
  [\href{https://arxiv.org/abs/1602.03410}{{\ttfamily 1602.03410}}].

\bibitem{Heisenberg:2016eld}
L.~Heisenberg, R.~Kase and S.~Tsujikawa, \emph{{Beyond generalized Proca
  theories}}, \href{https://doi.org/10.1016/j.physletb.2016.07.052}{\emph{Phys.
  Lett. B} {\bfseries 760} (2016) 617}
  [\href{https://arxiv.org/abs/1605.05565}{{\ttfamily 1605.05565}}].

\bibitem{Kimura:2016rzw}
R.~Kimura, A.~Naruko and D.~Yoshida, \emph{{Extended vector-tensor theories}},
  \href{https://doi.org/10.1088/1475-7516/2017/01/002}{\emph{JCAP} {\bfseries
  01} (2017) 002} [\href{https://arxiv.org/abs/1608.07066}{{\ttfamily
  1608.07066}}].

\bibitem{deRham:2020yet}
C.~de~Rham and V.~Pozsgay, \emph{{New class of Proca interactions}},
  \href{https://doi.org/10.1103/PhysRevD.102.083508}{\emph{Phys. Rev. D}
  {\bfseries 102} (2020) 083508}
  [\href{https://arxiv.org/abs/2003.13773}{{\ttfamily 2003.13773}}].

\bibitem{deRham:2021efp}
C.~de~Rham, S.~Garcia-Saenz, L.~Heisenberg and V.~Pozsgay, \emph{{Cosmology of
  Extended Proca-Nuevo}},  \href{https://arxiv.org/abs/2110.14327}{{\ttfamily
  2110.14327}}.

\bibitem{Aoki:2021wew}
K.~Aoki, M.~A. Gorji, S.~Mukohyama and K.~Takahashi, \emph{{The Effective Field
  Theory of Vector-Tensor Theories}},
  \href{https://arxiv.org/abs/2111.08119}{{\ttfamily 2111.08119}}.

\bibitem{Hull:2015uwa}
M.~Hull, K.~Koyama and G.~Tasinato, \emph{{Covariantized vector Galileons}},
  \href{https://doi.org/10.1103/PhysRevD.93.064012}{\emph{Phys. Rev.}
  {\bfseries D93} (2016) 064012}
  [\href{https://arxiv.org/abs/1510.07029}{{\ttfamily 1510.07029}}].

\bibitem{Horndeski:1974wa}
G.~W. Horndeski, \emph{{Second-order scalar-tensor field equations in a
  four-dimensional space}},
  \href{https://doi.org/10.1007/BF01807638}{\emph{Int. J. Theor. Phys.}
  {\bfseries 10} (1974) 363}.

\bibitem{Dias:2015nua}
O.~J.~C. Dias, J.~E. Santos and B.~Way, \emph{{Numerical Methods for Finding
  Stationary Gravitational Solutions}},
  \href{https://doi.org/10.1088/0264-9381/33/13/133001}{\emph{Class. Quant.
  Grav.} {\bfseries 33} (2016) 133001}
  [\href{https://arxiv.org/abs/1510.02804}{{\ttfamily 1510.02804}}].

\bibitem{Schutz:1985km}
B.~F. Schutz and C.~M. Will, \emph{{BLACK HOLE NORMAL MODES: A SEMIANALYTIC
  APPROACH}}, \href{https://doi.org/10.1086/184453}{\emph{Astrophys. J. Lett.}
  {\bfseries 291} (1985) L33}.

\bibitem{Horowitz:1999jd}
G.~T. Horowitz and V.~E. Hubeny, \emph{{Quasinormal modes of AdS black holes
  and the approach to thermal equilibrium}},
  \href{https://doi.org/10.1103/PhysRevD.62.024027}{\emph{Phys. Rev. D}
  {\bfseries 62} (2000) 024027}
  [\href{https://arxiv.org/abs/hep-th/9909056}{{\ttfamily hep-th/9909056}}].

\bibitem{Fang:2005qq}
H.~Fang and G.~Lovelace, \emph{{Tidal coupling of a Schwarzschild black hole
  and circularly orbiting moon}},
  \href{https://doi.org/10.1103/PhysRevD.72.124016}{\emph{Phys. Rev. D}
  {\bfseries 72} (2005) 124016}
  [\href{https://arxiv.org/abs/gr-qc/0505156}{{\ttfamily gr-qc/0505156}}].

\bibitem{Pani:2015hfa}
P.~Pani, L.~Gualtieri, A.~Maselli and V.~Ferrari, \emph{{Tidal deformations of
  a spinning compact object}},
  \href{https://doi.org/10.1103/PhysRevD.92.024010}{\emph{Phys. Rev. D}
  {\bfseries 92} (2015) 024010}
  [\href{https://arxiv.org/abs/1503.07365}{{\ttfamily 1503.07365}}].

\bibitem{Chakrabarti:2013lua}
S.~Chakrabarti, T.~Delsate and J.~Steinhoff, \emph{{New perspectives on neutron
  star and black hole spectroscopy and dynamic tides}},
  \href{https://arxiv.org/abs/1304.2228}{{\ttfamily 1304.2228}}.

\bibitem{Poisson:2020vap}
E.~Poisson, \emph{{Compact body in a tidal environment: New types of
  relativistic Love numbers, and a post-Newtonian operational definition for
  tidally induced multipole moments}},
  \href{https://doi.org/10.1103/PhysRevD.103.064023}{\emph{Phys. Rev. D}
  {\bfseries 103} (2021) 064023}
  [\href{https://arxiv.org/abs/2012.10184}{{\ttfamily 2012.10184}}].

\bibitem{Ramazanoglu:2017xbl}
F.~M. Ramazano\u{g}lu, \emph{{Spontaneous growth of vector fields in gravity}},
  \href{https://doi.org/10.1103/PhysRevD.96.064009}{\emph{Phys. Rev. D}
  {\bfseries 96} (2017) 064009}
  [\href{https://arxiv.org/abs/1706.01056}{{\ttfamily 1706.01056}}].

\bibitem{Ramazanoglu:2018tig}
F.~M. Ramazano\u{g}lu, \emph{{Spontaneous growth of gauge fields in gravity
  through the Higgs mechanism}},
  \href{https://doi.org/10.1103/PhysRevD.98.044013}{\emph{Phys. Rev. D}
  {\bfseries 98} (2018) 044013}
  [\href{https://arxiv.org/abs/1804.03158}{{\ttfamily 1804.03158}}].

\bibitem{Annulli:2019fzq}
L.~Annulli, V.~Cardoso and L.~Gualtieri, \emph{{Electromagnetism and hidden
  vector fields in modified gravity theories: spontaneous and induced
  vectorization}},
  \href{https://doi.org/10.1103/PhysRevD.99.044038}{\emph{Phys. Rev. D}
  {\bfseries 99} (2019) 044038}
  [\href{https://arxiv.org/abs/1901.02461}{{\ttfamily 1901.02461}}].

\bibitem{Minamitsuji:2020pak}
M.~Minamitsuji, \emph{{Spontaneous vectorization in the presence of vector
  field coupling to matter}},
  \href{https://doi.org/10.1103/PhysRevD.101.104044}{\emph{Phys. Rev. D}
  {\bfseries 101} (2020) 104044}
  [\href{https://arxiv.org/abs/2003.11885}{{\ttfamily 2003.11885}}].

\bibitem{Silva:2021jya}
H.~O. Silva, A.~Coates, F.~M. Ramazano\u{g}lu and T.~P. Sotiriou, \emph{{The
  ghost of vector fields in compact stars}},
  \href{https://arxiv.org/abs/2110.04594}{{\ttfamily 2110.04594}}.

\bibitem{Demirboga:2021nrc}
E.~S. Demirbo\u{g}a, A.~Coates and F.~M. Ramazano\u{g}lu, \emph{{Instability of
  vectorized stars}},  \href{https://arxiv.org/abs/2112.04269}{{\ttfamily
  2112.04269}}.

\bibitem{Berrut}
J.-P. Berrut and L.~N. Trefethen, \emph{Barycentric lagrange interpolation},
  \href{https://doi.org/10.1137/S0036144502417715}{\emph{SIAM Review}
  {\bfseries 46} (2004) 501}
  [\href{https://arxiv.org/abs/https://doi.org/10.1137/S0036144502417715}{{\ttfamily
  https://doi.org/10.1137/S0036144502417715}}].

\bibitem{Higham2004TheNS}
N.~Higham, \emph{The numerical stability of barycentric lagrange
  interpolation}, {\emph{Ima Journal of Numerical Analysis} {\bfseries 24}
  (2004) 547}.

\end{thebibliography}\endgroup

\end{document}